\documentclass[aps,prd,preprint,superscriptaddress,showpacs,nofootinbib]{revtex4}

\usepackage{graphicx}

\begin{document}

% Use the \preprint command to place your local institutional report
% number in the upper righthand corner of the title page in preprint mode.
% Multiple \preprint commands are allowed.
% Use the 'preprintnumbers' class option to override journal defaults
% to display numbers if necessary
\preprint{UTAP-419}

%Title of paper
\title{Three-generation study of neutrino spin-flavor conversion in
supernova and implication for neutrino magnetic moment}

% repeat the \author .. \affiliation  etc. as needed
% \email, \thanks, \homepage, \altaffiliation all apply to the current
% author. Explanatory text should go in the []'s, actual e-mail
% address or url should go in the {}'s for \email and \homepage.
% Please use the appropriate macro foreach each type of information

% \affiliation command applies to all authors since the last
% \affiliation command. The \affiliation command should follow the
% other information
% \affiliation can be followed by \email, \homepage, \thanks as well.
\author{Shin'ichiro Ando}
\email[E-mail: ]{ando@utap.phys.s.u-tokyo.ac.jp}
%\homepage[]{Your web page}
%\thanks{}
%\altaffiliation{}
\affiliation{Department of Physics, School of Science, the University of
Tokyo, 7-3-1 Hongo, Bunkyo-ku, Tokyo 113-0033, Japan}

\author{Katsuhiko Sato}
\affiliation{Department of Physics, School of Science, the University of
Tokyo, 7-3-1 Hongo, Bunkyo-ku, Tokyo 113-0033, Japan}
\affiliation{Research Center for the Early Universe, School of Science,
the University of Tokyo, 7-3-1 Hongo, Bunkyo-ku, Tokyo 113-0033, Japan}

%Collaboration name if desired (requires use of superscriptaddress
%option in \documentclass). \noaffiliation is required (may also be
%used with the \author command).
%\collaboration can be followed by \email, \homepage, \thanks as well.
%\collaboration{}
%\noaffiliation

\date{July 23, 2002; \today, revised}

\begin{abstract}
% insert abstract here
We investigate resonant spin-flavor (RSF) conversions of supernova
 neutrinos which are induced by the interaction of neutrino magnetic
 moment and supernova magnetic fields.
From the formulation which includes all three-flavor neutrinos and
 anti-neutrinos, we give a new crossing diagram that includes not only
 ordinary MSW resonance but also magnetically-induced RSF effect.
With the diagram, it is found that four conversions occur in supernova,
 two are induced by the RSF effect and two by the pure MSW.
We also numerically calculate neutrino conversions in supernova matter,
 using neutrino mixing parameters inferred from recent experimental
 results and a realistic supernova progenitor model.
The results indicate that until 0.5 seconds after core bounce, the
 RSF-induced $\bar{\nu}_e \leftrightarrow \nu_\tau$ transition occurs
 efficiently (adiabatic resonance), when $\mu_\nu \agt 10^{-12} \mu_B
 (B_0 / 5 \times 10^{9} \mathrm G)^{-1}$, where $B_0$ is the strength of
 the magnetic field at the surface of iron core.
We also evaluate the energy spectrum as a function of $\mu_\nu B_0$ at
 the SuperKamiokande detector and the Sudbury Neutrino Observatory using
 the calculated conversion probabilities, and find that the spectral
 deformation might have possibility to provide useful information on
 neutrino magnetic moment as well as magnetic field strength in
 supernovae.
\end{abstract}

% insert suggested PACS numbers in braces on next line
% insert suggested keywords - APS authors don't need to do this
%\keywords{}

%\maketitle must follow title, authors, abstract, \pacs, and \keywords
\maketitle

% body of paper here - Use proper section commands
% References should be done using the \cite, \ref, and \label commands
\section{Introduction \label{sec:Introduction}}
% Put \label in argument of %\section for cross-referencing
%\section{\label{}}
%\subsection{}
%\subsubsection{}

The properties of neutrinos attract a strong attention since neutrinos
alone are the elementary particles showing the evidence of new physics
beyond the standard model.
In particular, recent experiments of the SuperKamiokande (SK)
detector \cite{Totsuka92} and the Sudbury Neutrino Observatory (SNO)
\cite{Boger00} have shown that neutrino have non-zero mass and mixing
angles by solar \cite{Fukuda01,Bahcall98,Ahmad01,Ahmad02} and
atmospheric \cite{Fukuda99,Kajita01} neutrino data.
These neutrino oscillation experiments enable us to constrain the
squared mass difference between mass eigenstates and mixing angles.
The solar neutrino experiments strongly support the matter-induced large
mixing angle (LMA) solution ($\Delta m^2_{12}=2-10\times10^{-5}
~\mathrm{eV^2}, ~\tan^2\theta_{12}=0.25-0.5$ \cite{Ahmad02}) and the
atmospheric neutrino experiments have proved that $\nu_\mu$ and
$\nu_\tau$ are maximally mixing ($\Delta m^2_{23}=2-5\times10^{-3}
~\mathrm{eV^2}, ~\sin^22\theta_{23}>0.88$ \cite{Kajita01}).
However, there are several parameters which are not sufficiently
constrained. 
One such parameter is $\theta_{13}$, for which only upper bound is known
from reactor experiment \cite{Apollonio99} ($\sin^2 \theta_{13} \alt
0.1$).
In addition, the nature of neutrino mass hierarchy (normal or inverted)
is also matter of controversy.

There is another astrophysical source of neutrinos, or supernova.
Neutrino burst from the next Galactic core-collapse supernova will be
captured by the present underground detectors, such as SK and SNO.
It is widely believed that the detected events will provide rich
information not only on the mechanism of supernovae but also on the
physical properties of neutrinos themselves.
In fact, neutrino signals from SN1987A which were detected by the
Kamiokande \cite{Hirata87} and IMB \cite{Bionta87} detectors
permitted many discussions about the fundamental properties of
neutrinos \cite{Sato87,Arafune87,Goldman88}, although the detected
event numbers were quite small (11 for Kamiokande, 8 for IMB).
What we can learn from the next Galactic supernova is considered in many
articles (for a review, see Ref. \cite{Raffelt02}). 
For example, we can constrain the properties of neutrino oscillation
such as $\theta_{13}$ and the mass hierarchy \cite{Dighe00,Takahashi01}. 

In addition to neutrino mass, non-zero magnetic moment is another nature
of neutrinos beyond the standard model.
The most stringent upper bound of the neutrino magnetic moment is
obtained from cooling argument: any new physics that would enhance the
neutrino diffusion, such that the cooling time drops below the observed
duration, must be excluded.
The recently revised upper bound is $\mu_\nu \alt 1-4 \times
10^{-12} \mu_B$ \cite{Ayala99}, where $\mu_B$ is the Bohr magneton.
The above value is several orders of magnitude more stringent than the
existing limit from reactor \cite{Derbin93} and solar neutrino
experiments \cite{Gninenko99,Beacom99,Joshipura01}, and comparable to the 
bound from globular cluster red giant cooling arguments
\cite{Raffelt90}, which give $\mu_\nu \alt 3\times10^{-12}\mu_B$.
On the other hand, from the minimally-extended standard model
calculation, we obtain $\mu_\nu \simeq 3 \times 10^{-19} \mu_B (m_\nu /
1 ~\mathrm{eV})$, which is many orders of magnitude smaller than the
current experimental upper bounds.
However, some particle physics models have been proposed to induce a
larger magnetic moment of order $\sim 10^{-11} \mu_B$ \cite{Fukugita87},
which is comparable to the current upper bounds.

If neutrinos have the non-zero magnetic moment, it leads to precession
between left- and right-handed neutrinos in sufficiently strong magnetic
fields expected in the inner region of the core-collapse supernova
\cite{Cisneros70,Fujikawa80}.
In general, non-diagonal elements of the magnetic moment matrix are
possible and neutrinos can be changed into different flavors and
chiralities \cite{Schechter81}.
Furthermore, with the additional effect of the coherent forward
scattering by matter, neutrinos can be resonantly converted into those
with different chiralities \cite{Lim88} by a mechanism similar
to the well known Mikheyev-Smirnov-Wolfenstein (MSW) effect
\cite{Wolfenstein78,Kuo89}.
This resonant spin-flavor (RSF) conversion induced by the neutrino
magnetic moment in the supernova has been studied by many authors
\cite{Lim88,Akhmedov88,Voloshin88,Akhmedov92,Peltoniemi92,Athar95,
Totani96,Nunokawa97,Esposito97}. 
Almost all these earlier publications considered the RSF conversions
between two-flavor neutrinos ($\bar{\nu}_e \leftrightarrow \nu_\mu$).
However, since all three-flavor neutrinos and anti-neutrinos are
emitted from the supernova core, we should calculate conversion
probabilities using three-flavor formulation to obtain the spectrum of
the neutrinos in the detectors.
Fortunately, our knowledge of the neutrino mixing parameters permits
the three-flavor analyses, which have not been considered because of
many unknown parameters in those days.

In this paper, we study the RSF effect in detail, including three flavor
neutrinos and anti-neutrinos.
Particularly, a new crossing diagram, which deals with both MSW and RSF
effects, is given.
From the diagram, we find that the combination of the MSW and the RSF
effects makes crossing schemes very interesting to investigate, and the
expected flux would be different from that obtained in the case of pure
MSW or pure RSF effect.
In addition, we calculate numerically the conversion probabilities of
neutrinos and anti-neutrinos in the supernova using three-flavor
formulation. 
In the calculations, we adopt the latest neutrino mixing parameters,
which are constrained by the solar and the atmospheric neutrino
experiments, and assume the normal mass hierarchy. 
However, there are two unknown parameters of neutrinos, $\mu_\nu$ and
$\theta_{13}$ \footnote{There are many unknown properties in supernova
physics, such as density profile, $Y_e$, and magnetic field. We give in
Section \ref{sec:Models} the adopted properties in our calculation.}. 
For $\mu_\nu$, we assume the value $10^{-12}\mu_B$ \footnote{Since the
neutrino magnetic moment enters into the evolution equation only in the
combination $\mu_\nu B$ (see Section \ref{sec:Formulation}), the results
would also apply to any other value of $\mu_\nu$ provided that the
supernova magnetic field strength is rescaled accordingly.}, and for 
$\theta_{13}$, two large and small values ($\sin^2 2\theta_{13} = 0.04,
10^{-6}$). 
Using the calculated conversion probabilities, we give the expected
energy spectra at SK and SNO detectors, which show various behaviors as
we change the value of $\mu_\nu B$.
Finally, we discuss the model uncertainties and how to obtain the
implication for the value of neutrino magnetic moment from the obtained
energy spectra.

This paper is organized as follows.
In Section \ref{sec:Formulation}, we give the formulation used in our
calculation, which includes all three flavor neutrinos and
anti-neutrinos.
In Section \ref{sec:Models}, the neutrino oscillation models and
supernova models (original neutrino spectrum, density, $Y_e$, and
magnetic field profiles) considered in this paper are illustrated and
their validity is also discussed. 
In Section \ref{sec:Level crossing schemes and qualitative
illustrations}, we give a new crossing diagram, with which qualitative
discussions concerning neutrino conversions in supernova matter become
possible.
The numerically calculated conversion probabilities and energy spectra
at SK and SNO are shown in Section \ref{sec:Results}.
Finally, detailed discussions on model uncertainties and how to obtain
the implication for neutrino magnetic moment are presented in Section
\ref{sec:Discussion}.

% If in two-column mode, this environment will change to single-column
% format so that long equations can be displayed. Use
% sparingly.
%\begin{widetext}
% put long equation here
%\end{widetext}

% figures should be put into the text as floats.
% Use the graphics or graphicx packages (distributed with LaTeX2e)
% and the \includegraphics macro defined in those packages.
% See the LaTeX Graphics Companion by Michel Goosens, Sebastian Rahtz,
% and Frank Mittelbach for instance.

\section{Formulation \label{sec:Formulation}}

The interaction of the magnetic moment of neutrinos and magnetic fields
is described by
\begin{equation}
\langle(\nu_i)_R|H_{\mathrm{int}}|(\nu_j)_L \rangle  = \mu_{ij} B_\perp ,
\label{eq:interaction Hamiltonian}
\end{equation}
where $\mu_{ij}$ is the component of the neutrino magnetic moment
matrix, $B_\perp$ the magnetic field transverse to the direction of
propagation, $(\nu)_R$ and $(\nu)_L$ the right- and left-handed
neutrinos, respectively, and $i$ and $j$ denote the flavor eigenstate of
neutrinos, i.e., $e,~\mu,$ and $\tau$.
If neutrinos are the Dirac particles, right-handed neutrinos and
left-handed anti-neutrinos are undetectable (sterile neutrinos), since
they do not interact with matter.
The spin-flavor conversion into these sterile neutrinos suffers strong
constraints from observation of neutrinos from SN1987A
\cite{Voloshin88,Peltoniemi92,Lattimer88}.
On the other hand, if neutrinos are the Majorana particles, $\nu_R$'s
are identical to anti-particles of $\nu_L$'s and interact with matter,
and the constraint above becomes considerably weak.
In this paper, we assume that neutrinos are Majorana particles.
The diagonal magnetic moments is forbidden for the Majorana neutrinos,
and therefore, only the conversion between different flavors is
possible, e.g., $(\bar{\nu}_e)_R \leftrightarrow (\nu_{\mu,\tau})_L$.

The coherent forward scattering with matter induces the effective
potential for neutrinos, which is calculated using weak interaction
theory.
The effective potential due to scattering with electrons is given as
\begin{equation}
V_{\pm\pm} = \pm \sqrt{2} G_F \left( \pm \frac{1}{2} 
			       + 2\sin^2 \theta_W \right) n_e, 
\label{eq:electron potential}
\end{equation}
where $n_e$ is electron number density, $G_F$ the Fermi coupling
constant, and $\theta_W$ the Weinberg angle.
The $\pm$ sign in front refers to $\nu~(+)$ and $\bar{\nu}~(-)$ and that
in the parentheses to $\nu_e~(+)$ and $\nu_{\mu,\tau}~(-)$.
The difference between $e$ and $\mu,\tau$ neutrinos comes from the
existence of charged-current interaction.
The subscript $\pm\pm$ of $V$ refers to the first and the second $\pm$
sign.
The ordinary MSW effect between $\nu_e$ and $\nu_{\mu,\tau}$ is caused by
the potential difference, $V_e-V_{\mu,\tau} = V_{++}-V_{+-} = \sqrt{2}
G_F n_e$.
To include the RSF effect, which causes the conversion between neutrinos
and anti-neutrinos, we should take into account the neutral-current
scattering by nucleons:
\begin{equation}
V = \sqrt{2} G_F \left(\frac{1}{2} - 2 \sin^2 \theta_W \right) n_p 
 - \sqrt{2} G_F \frac{1}{2} n_n,
\label{eq:nucleon potential}
\end{equation}
where $n_p,~n_n$ are proton and neutron number density, respectively.
For neutrinos we add $+V$ to the potential and for anti-neutrinos $-V$.
Therefore, the RSF conversion between $\bar{\nu}_e$ and $\nu_{\mu,\tau}$
obeys the potential difference 
\begin{eqnarray}
\Delta V &\equiv& V_{\bar{e}}-V_{\mu,\tau} \nonumber \\
 &=& ( V_{-+} - V )  - ( V_{+-} + V )  \nonumber \\
 &=& \sqrt{2} G_F \frac{\rho}{m_N} ( 1 - 2 Y_e ),
\label{eq:Delta V}
\end{eqnarray}
where $\rho$ is the density, $m_N$ the nucleon mass, and $Y_e=n_e /
(n_e+n_n)$ is the number of electrons per baryon. (When we obtain
Eq. (\ref{eq:Delta V}), we assumed charge neutrality $n_e=n_p$.)

\subsection{The simplest case: $\bar{\nu}_e\leftrightarrow \nu_\mu$
conversion \label{sub:simplest}}

In this subsection, we give the simplest formulation between
$\bar{\nu}_e$ and $\nu_\mu$, and consider the properties of the RSF
conversion.
The time evolution of the mixed state of $\bar{\nu}_e$ and $\nu_\mu$ is
described by the Schr\"odinger equation
\begin{equation}
i \frac{d}{dr} \left( 
	      \begin{array}{c}
	      \bar{\nu}_e \\
	      \nu_\mu
	      \end{array}
	      \right)
= \left( 
   \begin{array}{cc}
    0 & \mu_{e\mu} B_\perp \\
    \mu_{e\mu} B_\perp & \Delta H
     \end{array}
 \right) \left( \begin{array}{c}
	\bar{\nu}_e \\
	\nu_\mu
	\end{array} \right),
\label{eq:two-flavor}
\end{equation}
where $r$ is the radius from the center of the star, $\mu_{e\mu}$ the
transition magnetic moment, and $\Delta H$ is defined by 
\begin{equation}
\Delta H \equiv \frac{\Delta m^2_{12}} 
{2E_\nu} \cos 2\theta_{12} - \Delta V.
\label{eq:Delta H}
\end{equation}
The resonance occurs when $\Delta H=0$, thus, the necessary condition is
$Y_e<0.5$.
(If $Y_e>0.5$, the $\nu_e \leftrightarrow \bar{\nu}_\mu $ conversion
occurs conversely, at almost the same place.)
Several earlier authors \cite{Athar95,Totani96} pointed out that the RSF
conversion occurs quite efficiently in the region above the iron core
and below the hydrogen envelope of collapsing stars, namely, in the
``isotopically neutral region'' (O+Si, O+Ne+Mg, O+C, and He layers), in
which the value of $Y_e$ is slightly less than 0.5 [typically, $(1-2Y_e)
\sim 10^{-4}-10^{-3}$]. 
Actually, the pre-supernova model used in our calculation have the
similar profile, as is shown in Section \ref{sub:Supernova progenitor
model}.

When the resonance is adiabatic (or when the density is slowly changing
at that point and the magnetic field is strong enough), the significant
conversion occurs. 
The adiabaticity of the RSF conversion can be derived by the same
argument used in the case of the MSW conversion \cite{Kuo89,Athar95}. 
It is given by
\begin{equation}
\gamma^{\mathrm{RSF}}_{\bar{e}\mu} = \frac{(2\mu_{e\mu}B_\perp)^2}
 {|d\Delta V / dr|}\biggm|_{\mathrm{res}},
\label{eq:adiabaticity}
\end{equation}
where the subscript ``res'' means that the value is evaluated at the
resonance point.
Therefore, if the magnetic field is sufficiently strong at the resonance
point, the $\bar{\nu}_e \leftrightarrow \nu_\mu$ conversion occurs
completely.

At the end of this subsection, we discuss the possibility that $\nu_e
\leftrightarrow \bar{\nu}_\mu$ conversion occurs in solar magnetic
field. 
There are several works examining the RSF conversion as a solution to
the solar neutrino problem \cite{Pulido00}.
In this paper, however, we postulate that the solution to the solar
neutrino problem is given by the ordinary MSW effect (LMA solution).
In fact, if we take the mixing parameters of LMA ($\Delta m^2_{12} = 2-10
\times 10^{-5}~\mathrm{eV^2},~\tan^2 \theta_{12} = 0.25-0.5$) in
calculating the RSF resonance condition $\Delta H = 0$, the required
density is found to be larger than the central density of the Sun.
(In this calculation, the sign of $\Delta V$ is changed in
Eq. (\ref{eq:Delta H}), since $Y_e>0.5$ in the Sun. This corresponds to
$\nu_e \leftrightarrow \bar{\nu}_\mu$ conversion.)

\subsection{Three-flavor formulation \label{sub:Three-flavor formulation}}

Here, we present the three-flavor (six-component) formulation of
neutrino mixing by
\begin{equation}
i \frac{d}{dr} \left(
	      \begin{array}{c} 
	       \nu \\
	       \bar{\nu}
	      \end{array}
	      \right)
= \left( \begin{array}{cc}
 H_0 & B_\perp M \\ -B_\perp M & \bar{H_0} \end{array} \right)
\left( \begin{array}{c} \nu \\ \bar{\nu} \end{array} \right),
\label{eq:three-flavor}
\end{equation}
where 
\begin{equation}
\nu = \left(
	 \begin{array}{c}
	  \nu_e \\
	  \nu_\mu \\
	  \nu_\tau
	 \end{array}
       \right), 
~~\bar{\nu} = \left(
	 \begin{array}{c}
	  \bar{\nu}_e \\
	  \bar{\nu}_\mu \\
	  \bar{\nu}_\tau
	 \end{array}
       \right), 
\end{equation}
\begin{widetext}
%\begin{subequations}
 \begin{eqnarray}
  H_0 = \frac{1}{2E_\nu} U \left(
			  \begin{array}{ccc}
			  0 & 0 & 0 \\
			  0 & \Delta m^2_{12} & 0 \\
			  0 & 0 & \Delta m^2_{13}
			  \end{array}
			  \right) U^\dagger
  + \left(
   \begin{array}{ccc}
   V_{++}+V & 0 & 0 \\
   0 & V_{+-}+V & 0 \\
   0 & 0 & V_{+-}+V 
   \end{array}
   \right), \label{equationa} 
\\
  \bar{H_0} = \frac{1}{2E_\nu} U \left(
			  \begin{array}{ccc}
			  0 & 0 & 0 \\
			  0 & \Delta m^2_{12} & 0 \\
			  0 & 0 & \Delta m^2_{13}
			  \end{array}
			  \right) U^\dagger
  + \left(
   \begin{array}{ccc}
   V_{-+}-V & 0 & 0 \\
   0 & V_{--}-V & 0 \\
   0 & 0 & V_{--}-V 
   \end{array}
   \right), \label{equationb}
\\
  U = \left(
       \begin{array}{ccc}
	U_{e1} & U_{e2} & U_{e3} \\
        U_{\mu 1} & U_{\mu 2} & U_{\mu 3} \\
        U_{\tau 1} & U_{\tau 2} & U_{\tau 3}
       \end{array}
     \right)
  = \left(
   \begin{array}{ccc}
   c_{12}c_{13} 
    & s_{12}c_{13} 
   & s_{13} \\
   -s_{12}c_{23}
    -c_{12}s_{23}s_{13}
  & c_{12}c_{23}
  -s_{12}s_{23}s_{13} 
  & s_{23}c_{13} \\
    s_{12}s_{23}
     -c_{12}c_{23}s_{13}
     & -c_{12}s_{23}
     -s_{12}c_{23}s_{13}
     & c_{23}c_{13}
     \end{array}
   \right), \label{eq:U}
 \end{eqnarray}
%\end{subequations}
\end{widetext}
\begin{equation}
M = \left(
   \begin{array}{ccc}
   0 & \mu_{e\mu} & \mu_{e\tau} \\
   -\mu_{e\mu} & 0 & \mu_{\mu\tau} \\
   -\mu_{e\tau} & -\mu_{\mu\tau} & 0
   \end{array}
   \right),
\label{eq:magnetic moment}
\end{equation}
and $c_{ij}=\cos\theta_{ij},~s_{ij}=\sin\theta_{ij}$.
(We assume CP phase $\delta=0$ in Eq. (\ref{eq:U}) for simplicity.)

The resonant flavor conversion occurs when the two diagonal elements in
matrix in Eq. (\ref{eq:three-flavor}) have the same value.
There are five such resonance points, which are for $\nu_e
\leftrightarrow \nu_\mu$ (MSW-L), $\nu_e \leftrightarrow \nu_\tau$
(MSW-H), $\bar{\nu}_e \leftrightarrow \nu_\mu$ (RSF-L), $\bar{\nu}_e
\leftrightarrow \nu_\tau$ (RSF-H), and $\bar{\nu}_\mu \leftrightarrow
\nu_\tau$ conversions.
Suffixes ``-L'' and ``-H'' attached to ``MSW'' and ``RSF'' indicate
whether the density at the resonance points are lower or higher.
Hereafter, we neglect $\bar{\nu}_\mu \leftrightarrow \nu_\tau$
conversion, since it is always nonadiabatic as well as including this
makes discussion further complicated.

\section{Models \label{sec:Models}}

\subsection{Neutrino parameters \label{sub:Neutrino parameters}}

We adopt the neutrino mixing parameters introduced to explain the
observations of the solar and the atmospheric neutrino experiments
\cite{Fukuda01,Bahcall98,Ahmad01,Ahmad02,Fukuda99,Kajita01},
and assume the normal mass hierarchy.
From the atmospheric neutrino experiments, we use the values
\cite{Kajita01}
\begin{equation}
\Delta m^2_{13} = 2.8 \times 10^{-3} ~\mathrm{eV^2},
~~\sin^22\theta_{23}=1.0,
\end{equation}
and from the solar neutrino experiments \cite{Ahmad02},
\begin{equation}
\Delta m^2_{12} = 5.0 \times 10^{-5} ~\mathrm{eV^2},
~~\tan^2 \theta_{12} = 0.34,
\end{equation}
in our calculations.
For $\theta_{13}$, which is not sufficiently constrained, we adopt two
large and small values,
\begin{equation}
\sin^2 2\theta_{13} = 0.04, ~1.0 \times 10^{-6},
\end{equation}
where each value causes the adiabatic and nonadiabatic MSW-H
conversions, respectively.

As for neutrino magnetic moment, there are three transition moments for
Majorana neutrino (Eq. (\ref{eq:magnetic moment})).
We assume that all of them are near the present upper bound (i.e.,
$\mu_{ij} = 10^{-12} \mu_B$).
However, as we will show in Section \ref{sub:Supernova magnetic field},
$\mu_{e\tau}$ alone governs results.

\subsection{Original spectrum of neutrinos \label{sub:Original spectrum
of neutrinos}}

\begin{figure}[htbp]
\includegraphics[width=15cm]{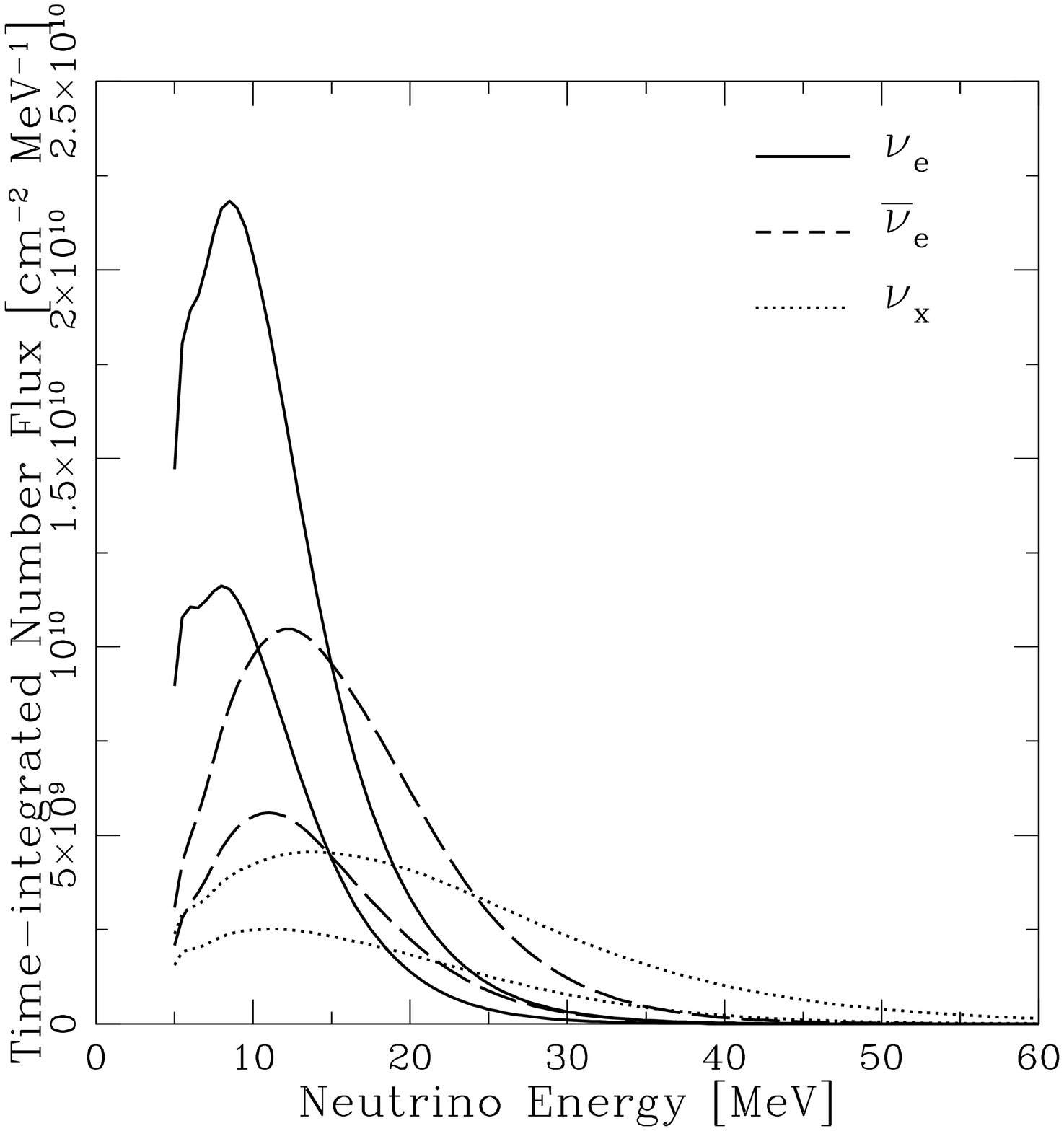}%
\caption{Original time-integrated energy spectra of $\nu_e$ (solid curve),
 $\bar{\nu}_e$ (dashed curve), and $\nu_x$ (dotted curve), where $\nu_x$
 represents $\nu_{\mu,\tau}$ and $\bar{\nu}_{\mu,\tau}$. For each line
 type, the larger one represents fully time-integrated spectrum, while
 smaller one spectrum integrated until 0.5 seconds after core
 bounce. \label{fig:fluence}}
\end{figure}

We use a realistic model of a collapse-driven supernova by the Lawrence
Livermore group \cite{Wilson86} to calculate the neutrino energy
spectrum at production.
We show in Fig. \ref{fig:fluence} the time-integrated energy spectra of
neutrinos (see Ref. \cite{Totani98} for detail), where the distance to
the supernova is set to 10 kpc. 
(For discussions from here, we adopt this assumption, $D=10
~\mathrm{kpc}$.)
Both the flux integrated over entire time during neutrino burst and that
integrated until 0.5 seconds after core bounce are shown in the same
figure. 
Although the neutrino burst persists $\sim$ 10 seconds, about half of
neutrinos are emitted by 0.5 seconds, according to numerical simulation
by Lawrence Livermore group \cite{Totani98}.
The reason for using time integrated flux until 0.5 seconds is given in
the next subsection.

The mean energies are different between flavors:
\begin{eqnarray}
\langle E_{\nu_e} \rangle &\simeq& 11 ~\mathrm{MeV}, \\
\langle E_{\bar{\nu}_e} \rangle &\simeq& 16 ~\mathrm{MeV}, \\
\langle E_{\nu_x} \rangle &\simeq& 22 ~\mathrm{MeV},
\end{eqnarray}
for fully time-integrated spectrum, while for that integrated until 0.5
seconds,
\begin{eqnarray}
\langle E_{\nu_e} \rangle &\simeq& 11 ~\mathrm{MeV}, \\
\langle E_{\bar{\nu}_e} \rangle &\simeq& 14 ~\mathrm{MeV}, \\
\langle E_{\nu_x} \rangle &\simeq& 18 ~\mathrm{MeV}.
\end{eqnarray}
This hierarchy of mean energies is explained as follows.
Since $\nu_x$'s interact with matter only through the neutral-current
reactions in supernova, they are weakly coupled with matter compared to
$\nu_e$'s and $\bar{\nu}_e$'s.
Thus the neutrino sphere of $\nu_x$'s is deeper in the core than that of
$\nu_e$'s and $\bar{\nu}_e$'s, which leads to higher temperatures for
$\nu_x$'s. 
The difference between $\nu_e$'s and $\bar{\nu}_e$'s comes from that the
core is neutron rich and $\nu_e$'s couples with matter more strongly,
through $\nu_e n \rightarrow e^- p$ reactions.

\subsection{Supernova progenitor model \label{sub:Supernova progenitor
model}}

We use the precollapse model of massive stars of Woosley and Weaver
\cite{Woosley95}.
The progenitor mass was set to be $15M_\odot$, and the metalicity the
same as that of the Sun.
The upper panel of Fig. \ref{fig:profiles} shows $\rho Y_e$ and $\rho
(1-2Y_e)$ profiles of this progenitor model.
In the same figure, we also show $\Delta_{12}\equiv \Delta m^2_{12} \cos
2 \theta_{12}/2 \sqrt{2}G_F E_\nu$ and $\Delta_{13} \equiv \Delta
m^2_{13} \cos 2 \theta_{13}/2 \sqrt{2} G_F E_\nu$.
At intersections between $\Delta_{12}, \Delta_{13}$ and $\rho (1-2Y_e),
\rho Y_e$, the RSF and MSW conversions occur (see Eq. (\ref{eq:Delta
H})).

However, in fact, since the density profile changes drastically during
neutrino burst ($\sim 10$ sec) owing to shock propagation, we cannot use
the static progenitor model when calculating the conversion probability
in supernova \cite{Schirato02}.
We show in Fig. \ref{fig:shock} the {\it time-dependent} $\rho (1-2Y_e)$
and $\rho Y_e$ profile calculated by Lawrence Livermore group
\cite{Takahashi02}, compared with the static model adopted in our
calculation \cite{Woosley95}.
From the figure, the shock wave changes the $\rho (1-2Y_e)$ profile in
the resonance region, which is responsible for RSF conversions, at
$> 0.5$ seconds after bounce, while that at $\alt 0.5$ seconds is
almost the same as the static progenitor model.
Thus, from here, we confine our discussion until 0.5 seconds after core
bounce, since in that case, using the static progenitor model is
considered to be a good approximation.
Calculating for full time-scale during neutrino burst with the
time-dependent density and $Y_e$ profiles makes discussion very
complicated, and it is out of scope of this study, while it will be our
future work.
%On the other hand, considering pure MSW effect with time-dependent
%profile is now in preparation \cite{Takahashi02}.
\subsection{Supernova magnetic field \label{sub:Supernova magnetic
field}}

We assume that the global structure of the magnetic field is a dipole
moment and the field strength is normalized at the surface of the iron
core with the value of $10^8-10^{10}$ Gauss. 
The reason for this normalization is as follows.
As pointed out by Totani and Sato \cite{Totani96}, the global magnetic
field normalized at the surface of the newly born neutron star, such as
used in Ref. \cite{Athar95}, is not a good assumption.
It is because that the magnetic fields of a nascent neutron star are
hard to affect the far outer region, such as the isotopically neutral
region, within the short time scale of neutrino burst.
The magnetic fields, therefore, should be normalized by the fields which
are static and existent before the core collapse.
In addition, as discussed in the previous subsection, since the shock
wave does not affect the resonance region at $\alt 0.5$ seconds after
bounce, it is also expected that the magnetic field strength at the
resonance points is not seriously changed at that time.
The strength of such magnetic fields above the surface of the iron core
may be inferred from the observation of the surface of white dwarfs,
since both are sustained against the gravitational collapse by the
degenerate pressure of electrons.
The observations of the magnetic fields in white dwarfs show that the
strength spreads in a wide range of $10^7-10^9$ Gauss
\cite{Chanmugam92}.
Taking account of the possibility of the decay of magnetic fields in
white dwarfs, it is not unnatural to consider the magnetic fields up to
$10^{10}$ Gauss at the surface of the iron core.
Then, in Eq. (\ref{eq:three-flavor}), $B_\perp = B_0 (r_0 / r)^3 \sin
\Theta$, where $B_0$ is the strength of the magnetic field at the
equator on the iron core surface, $r_0$ the radius of the iron core, and
$\Theta$ the angle between the pole of the magnetic dipole and the
direction of neutrino propagation. 
Hereafter, we assume $\sin \Theta = 1$.

From the magnetic field, density, and $Y_e$ profiles of supernova we can
calculate the adiabaticity parameter of each resonance points with
Eq. (\ref{eq:adiabaticity}).
The lower panel of Fig. \ref{fig:profiles} shows adiabaticity parameters
of RSF conversions as a function of radius.
Those for both $B_0=10^8$ and $B_0=10^{10}$ Gauss, when $\mu_\nu =
10^{-12} \mu_B$ is assumed, are shown.
From the figure, we see that the RSF-H conversion can be adiabatic when
the $B_0 \sim 10^{10}~\mathrm G$, while the RSF-L conversion is always
nonadiabatic under any conditions considered here.
Therefore, the values of transition magnetic moment $\mu_{e\mu},
\mu_{\mu\tau}$ do not matter, while that of $\mu_{e\tau}$ alone governs
results.

\section{Level crossing schemes and qualitative illustrations
\label{sec:Level crossing schemes and qualitative illustrations}}

\begin{figure}[htbp]
\includegraphics[width=15cm]{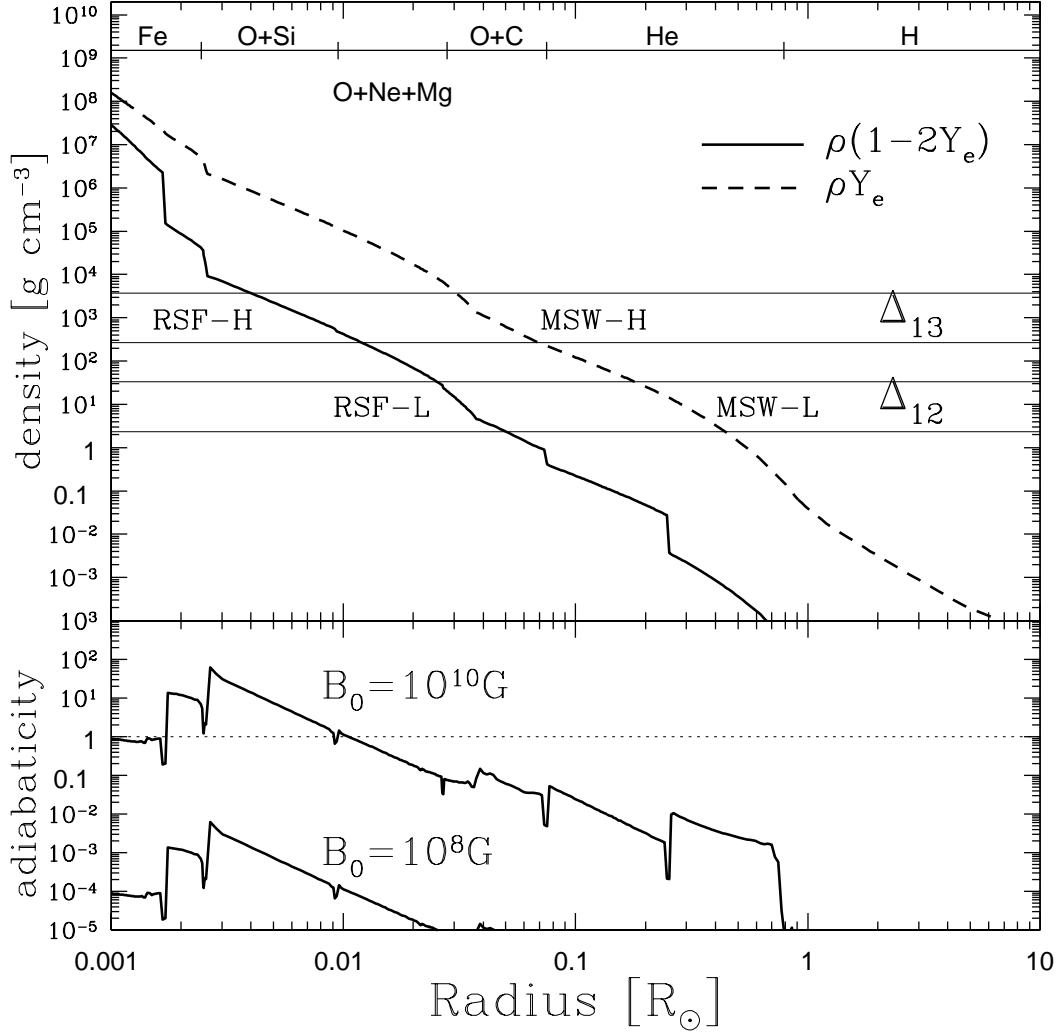}%
\caption{Supernova profiles used in our calculations
 \cite{Woosley95}. Upper panel: The solid curve represents the density
 and $Y_e$ combination which is responsible for the RSF conversions
 $(\rho (1-2Y_e))$, and the dashed curve for the MSW conversions
 $(\rho Y_e)$. Two horizontal bands are  for $\Delta_{12}=\Delta m^2_{12}
 \cos2 \theta_{12}/2 \sqrt{2} G_F E_\nu$ and $\Delta_{13} = \Delta
 m^2_{13} \cos2 \theta_{13} / 2 \sqrt{2} G_F E_\nu$. (The band width
 comes from the energy range 5-70 MeV.) At intersections between the
 solid, dashed curves and the horizontal bands, the RSF and MSW
 conversions occur (see Eq. (\ref{eq:Delta H})). Lower panel: The
 adiabaticity parameter of RSF conversions (Eq. (\ref{eq:adiabaticity}))
 as a function of radius, where the indicated values are magnetic field
 strength at the iron core ($B_0$) when $\mu_\nu = 10^{-12} \mu_B$ is
 assumed. \label{fig:profiles}}
\end{figure}

\begin{figure}[htbp]
\includegraphics[width=15cm]{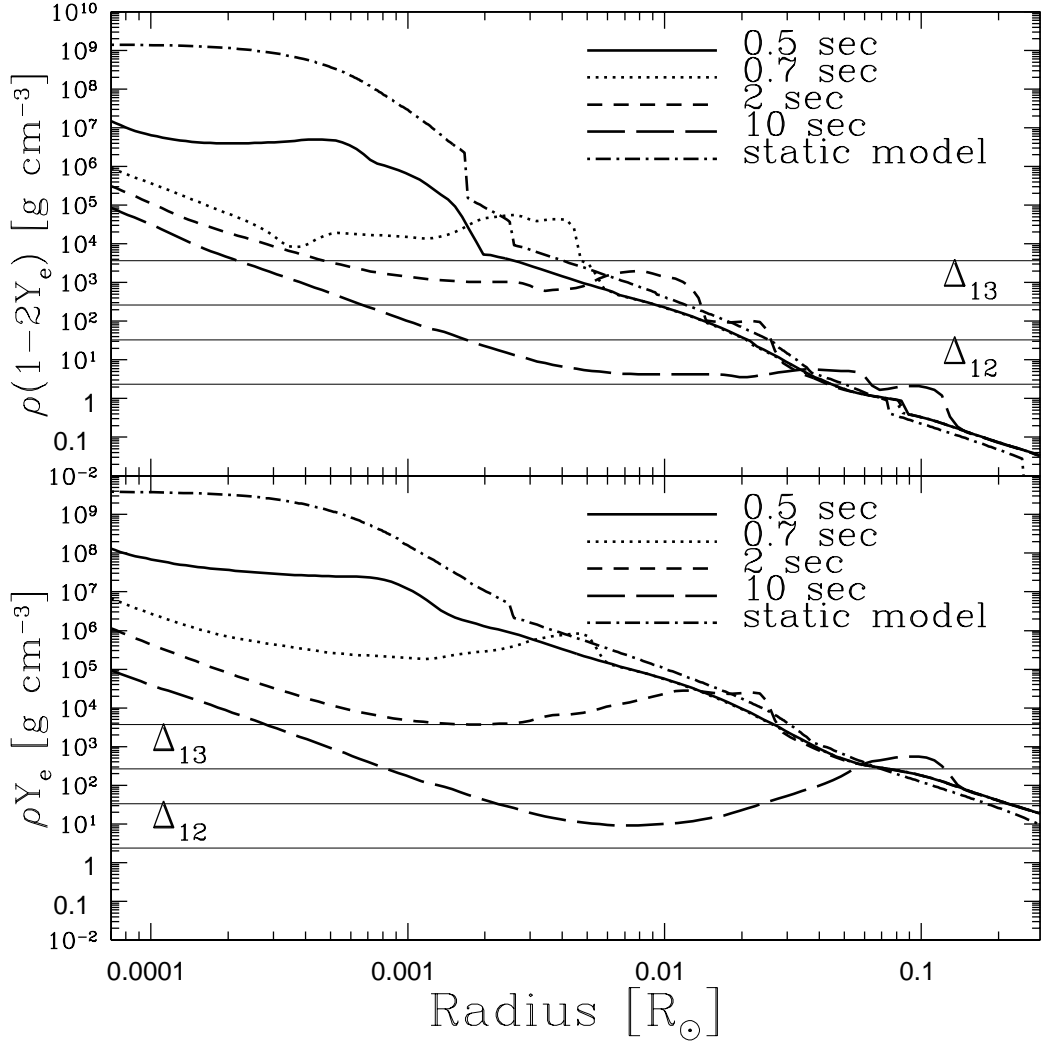}
\caption{Time-dependent density and $Y_e$ profiles of supernova
 calculated by Lawrence Livermore group \cite{Takahashi02}. Upper panel:
 The $\rho (1-2Y_e)$ profile of supernova, which is responsible for RSF
 conversions. The profile at 0.5, 0.7, 2, and 10 seconds after bounce,
 are shown. The profile calculated with static progenitor model
 \cite{Woosley95} is also shown, for comparison. The resonance points,
 which are represented by the intersections between the $\rho (1-2Y_e)$
 curves and horizontal bands ($\Delta_{13},\Delta_{12}$), are affected
 by the shock propagation at $> 0.5$ seconds. Lower panel: The same
 figure, but for $\rho Y_e$ profile, which is responsible for MSW
 conversions. \label{fig:shock}}
\end{figure}

\begin{figure}[htbp]
\includegraphics[width=15cm]{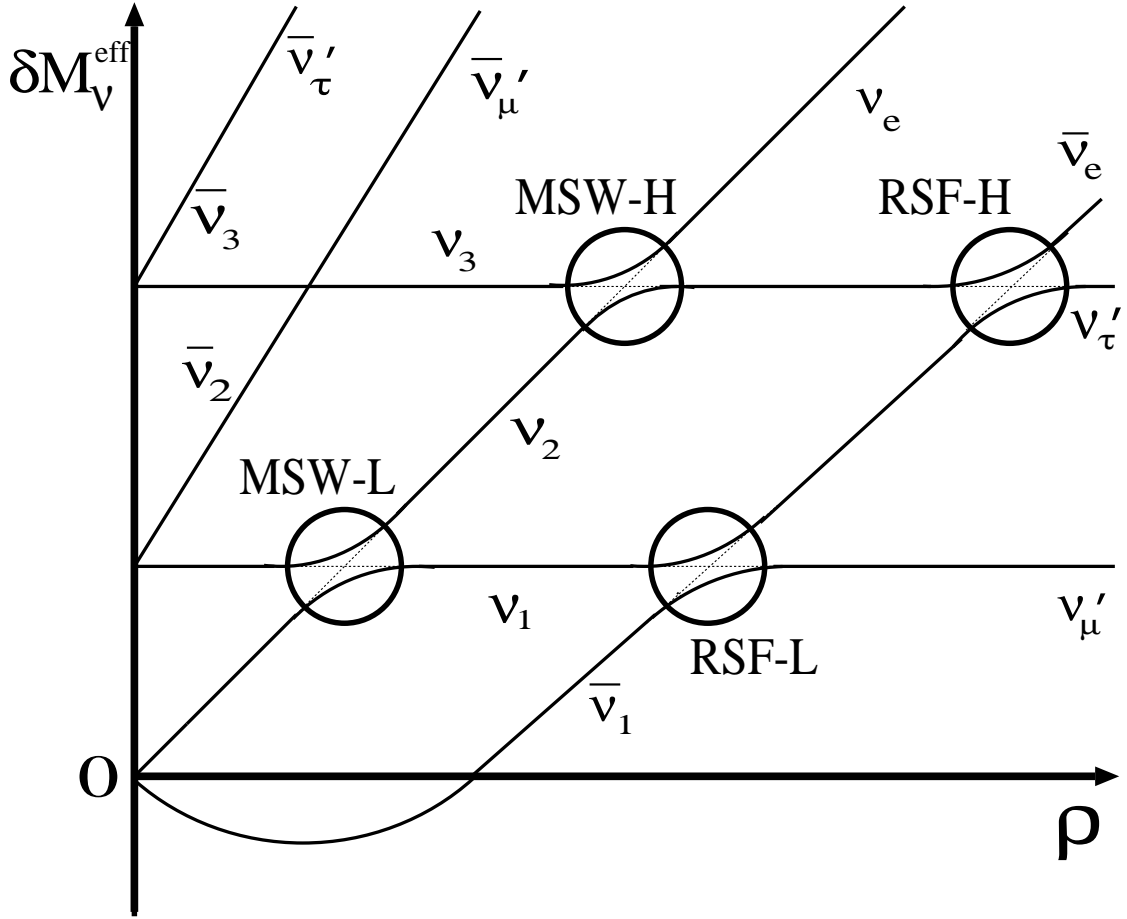}%
\caption{The schematic illustration of level crossings, where
 $\nu_{1,2,3}$ and $\bar{\nu}_{1,2,3}$ represent the mass eigenstates of
 neutrinos and anti-neutrinos in matter, respectively, and
 $\nu_{\mu,\tau}^\prime$ and $\bar{\nu}_{\mu,\tau}^\prime$ the mass
 eigenstates at production, which are superpositions of $\nu_\mu$ and
 $\nu_\tau$ or $\bar{\nu}_\mu$ and $\bar{\nu}_\tau$. There are four
 resonance points, MSW-L, MSW-H, RSF-L, and RSF-H. The adiabatic
 conversion means that the neutrinos trace the solid curve at each
 resonance point (i.e., the mass eigenstate does not flip), while the
 nonadiabatic conversion the dotted line. \label{fig:crossing}} 
\end{figure}

From the formulation which includes three-flavor neutrinos and
anti-neutrinos (Eq. (\ref{eq:three-flavor})), we give a new level
crossing diagram, which is an expanded version of the ordinary diagram
including MSW resonance alone (see Fig. 5 in Ref. \cite{Dighe00}).
We show it in Fig. \ref{fig:crossing}, in which the adiabatic conversion
means that the neutrinos trace the solid curve at each resonance point
(i.e., the mass eigenstate does not flip), while the nonadiabatic
conversion the dotted line.
The figure clearly includes not only ordinary MSW resonances but also
the RSF effects, and it is expected that the combination effect of MSW
and RSF makes this scheme very interesting to investigate.
From the earlier publications on the MSW effect, we know that the MSW-L
is adiabatic for LMA solution, and MSW-H is adiabatic for large
$\theta_{13}$ and nonadiabatic for small $\theta_{13}$
\cite{Dighe00,Takahashi01}.
The adiabaticity of the RSF-L and RSF-H depends on the magnetic field
strength and the density profile at the resonance point (see
Eq. (\ref{eq:adiabaticity})), which are given in Section
\ref{sec:Models} and Fig. \ref{fig:profiles}.

Here, we qualitatively consider the flavor transition of neutrinos under
some simple conditions.
At the supernova core, electron neutrinos and anti-neutrinos are
produced as the mass eigenstate in matter, owing to large matter
potential.
The other flavor ($\mu,\tau$) neutrinos and anti-neutrinos at
production are not mass eigenstates, however, we can make them by linear
combination of these flavor eigenstates.
(These states corresponds to $\nu_\mu^\prime, \nu_\tau^\prime,
\bar{\nu}_\mu^\prime, \bar{\nu}_\tau^\prime$ in Fig. \ref{fig:crossing}.)
Then, neutrinos and anti-neutrinos of all flavors propagate in supernova
matter along the curves in Fig. \ref{fig:crossing}, being affected by
four resonances, and appear from the stellar surface as pure vacuum
mass eigenstates or their mixing states.

When both the RSF-L and RSF-H are nonadiabatic, the situation is the
same as the pure MSW case, i.e., the conversions occurs as
\begin{eqnarray}
\begin{array}{cc}
\nu_e \rightarrow \nu_3 & \bar{\nu}_e \rightarrow \bar{\nu}_1 \\
\nu_\mu^\prime \rightarrow \nu_1 
& \bar{\nu}_\mu^\prime \rightarrow \bar{\nu}_2 \\
\nu_\tau^\prime \rightarrow \nu_2 
& \bar{\nu}_\tau^\prime \rightarrow \bar{\nu}_3
\end{array} &\Biggr\}& ~\mbox{(adiabatic MSW-H)}, \nonumber \\
\begin{array}{cc}
\nu_e \rightarrow \nu_2 & \bar{\nu}_e \rightarrow \bar{\nu}_1 \\
\nu_\mu^\prime \rightarrow \nu_1 
& \bar{\nu}_\mu^\prime \rightarrow \bar{\nu}_2 \\
\nu_\tau^\prime \rightarrow \nu_3 
& \bar{\nu}_\tau^\prime \rightarrow \bar{\nu}_3
\end{array} &\Biggr\}& ~\mbox{(nonadiabatic MSW-H)}. \nonumber 
\end{eqnarray}
From these considerations, we can estimate the spectral deformation of
$\nu_e$ and $\bar{\nu}_e$, which are the easiest to detect at SK and
SNO.
When the MSW-H is adiabatic, we obtain
\begin{eqnarray}
F_e &=& |U_{e1}|^2F_1 + |U_{e2}|^2F_2 + |U_{e3}|^2F_3 \nonumber \\
    &=& |U_{e1}|^2F_x^0 + |U_{e2}|^2F_x^0 + |U_{e3}|^2F_e^0 \nonumber \\
    &=& (1-|U_{e3}|^2)F_x^0 + |U_{e3}|F_e^0, 
\label{eq:F_e_A_NA} \\
F_{\bar{e}} &=& |U_{e1}|^2F_{\bar{e}}^0 + (1-|U_{e1}|^2)F_x^0,
\label{eq:F_ebar_A_NA}
\end{eqnarray}
where $F_i$ is the flux of the $i$-type neutrinos at detection, $F_i^0$
the flux at production which is shown in Fig. \ref{fig:fluence}, and
subscript $x = \mu, \tau, \bar{\mu}, \bar{\tau}$, since the flux of
these flavors are considered to be the same.
With the same manner, for nonadiabatic MSW-H, we obtain
\begin{eqnarray}
F_e &=& (1-|U_{e2}|^2)F_x^0 + |U_{e2}|^2F_e^0, 
\label{eq:F_e_NA_NA} \\
F_{\bar{e}} &=& |U_{e1}|^2F_{\bar{e}}^0 + (1-|U_{e1}|^2)F_x^0.
\label{eq:F_ebar_NA_NA}
\end{eqnarray}

When the RSF-H is adiabatic and RSF-L nonadiabatic, which is actually
the case for strong magnetic field (see Fig. \ref{fig:profiles}), the
same arguments lead to for adiabatic MSW-H:
\[
\left\{
\begin{array}{cc}
\nu_e \rightarrow \nu_3 & \bar{\nu}_e \rightarrow \nu_2 \\
\nu_\mu^\prime \rightarrow \nu_1 
& \bar{\nu}_\mu^\prime \rightarrow \bar{\nu}_2 \\
\nu_\tau^\prime \rightarrow \bar{\nu}_1 
& \bar{\nu}_\tau^\prime \rightarrow \bar{\nu}_3 \nonumber
\end{array}
\right. , \\
\]
\begin{eqnarray}
F_e &=& |U_{e1}|^2F_x^0 + |U_{e2}|^2F_{\bar{e}}^0 + |U_{e3}|^2F_e^0, 
\label{eq:F_e_A_A}\\
F_{\bar{e}} &=& F_x^0,
\label{eq:F_ebar_A_A}
\end{eqnarray}
and for nonadiabatic MSW-H:
\[
\left\{
\begin{array}{cc}
\nu_e \rightarrow \nu_2 & \bar{\nu}_e \rightarrow \nu_3 \\
\nu_\mu^\prime \rightarrow \nu_1 
& \bar{\nu}_\mu^\prime \rightarrow \bar{\nu}_2 \\
\nu_\tau^\prime \rightarrow \bar{\nu}_1 
& \bar{\nu}_\tau^\prime \rightarrow \bar{\nu}_3 \nonumber 
\end{array}
\right. , \\
\]
\begin{eqnarray}
F_e &=& |U_{e1}|^2F_x^0 + |U_{e2}|^2F_e^0 + |U_{e3}|^2F_{\bar{e}}^0, 
\label{eq:F_e_NA_A} \\
F_{\bar{e}} &=& F_x^0.
\label{eq:F_ebar_NA_A}
\end{eqnarray}
The adiabatic RSF-H conversion, thus, deforms the energy spectra, and
the difference from the case of the pure MSW effect is clear.

\section{Results of numerical calculations \label{sec:Results}}

\subsection{Conversion probability \label{sub:Conversion probability}}

\begin{figure}[htbp]
\includegraphics[width=15cm]{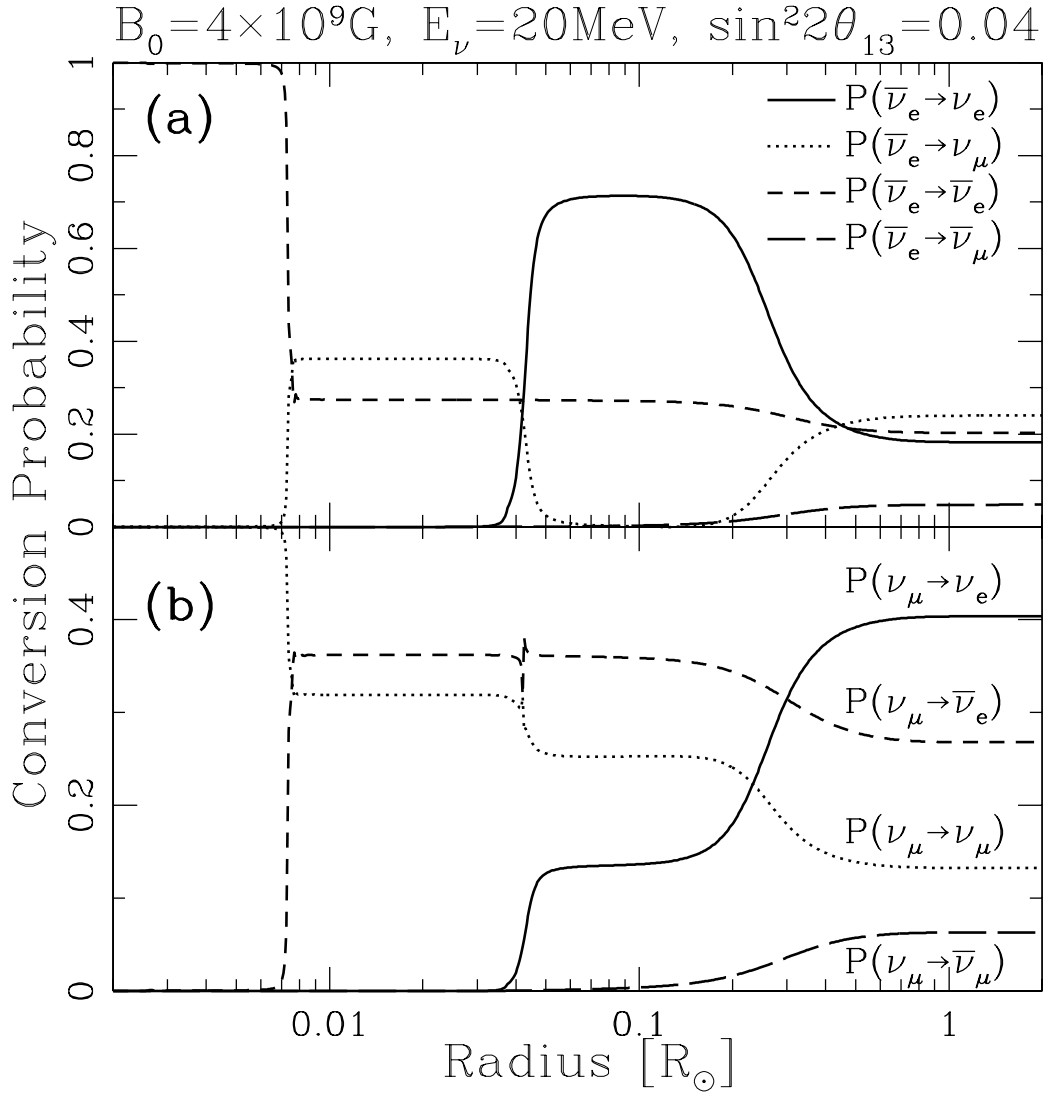}%
\caption{(a) Conversion probabilities of $\bar{\nu}_e$ under the
 condition that $B_0 = 4  \times 10^9 ~\mathrm G, ~E_\nu = 20
 ~\mathrm{MeV}$, and $\sin^2 2  \theta_{13} = 0.04$. The solid curve
 represents $P(\bar{\nu}_e \rightarrow \nu_e)$, the dotted curve
 $P(\bar{\nu}_e \rightarrow \nu_\mu)$, the short-dashed curve
 $P(\bar{\nu}_e \rightarrow \bar{\nu}_e)$, and the long-dashed curve
 $P(\bar{\nu}_e \rightarrow \bar{\nu}_\mu)$. (b) The same figure
 as (a), but for those of $\nu_\mu$. The solid curve represents
 $P(\nu_\mu \rightarrow \nu_e)$, the dotted curve $P(\nu_\mu \rightarrow
 \nu_\mu)$, the short-dashed curve $P(\nu_\mu \rightarrow \bar{\nu}_e)$,
 and the long-dashed curve $P(\nu_\mu \rightarrow \bar{\nu}_\mu)$.
 \label{fig:prob_L}}
\end{figure}

We calculated Eq. (\ref{eq:three-flavor}) numerically with adopted
models given in Section \ref{sec:Models}, and obtained the conversion
probabilities for each flavor.
Figure \ref{fig:prob_L} shows those for $\bar{\nu}_e$ (upper panel (a))
and $\nu_\mu$ (lower panel (b)), under the condition that $B_0 = 4
\times 10^9 ~\mathrm G, ~E_\nu = 20 ~\mathrm{MeV}$, and $\sin^2 2
\theta_{13} = 0.04$.
Since $\nu_\tau$'s ($\bar{\nu}_\tau$'s) are maximally mixing with
$\nu_\mu$'s ($\bar{\nu}_\mu$'s), we did not show $P(\bar{\nu}_e
\rightarrow \nu_\tau,\bar{\nu}_\tau)$ and $P(\nu_\mu \rightarrow
\nu_\tau,\bar{\nu}_\tau)$, which are almost the same as that for
$\nu_\mu,\bar{\nu}_\mu$.
We can see three clear conversions, one RSF and two MSWs.
The RSF-H conversion occurs at $r \sim 0.007 R_\odot$ between
$\bar{\nu}_e$ and $\nu_\tau$.
(The apparent $\bar{\nu}_e \leftrightarrow \nu_\mu$ conversion occurs,
since $\nu_\mu$'s are maximally mixing with $\nu_\tau$'s.)
After that, neutrinos propagate in matter inducing the ordinary MSW
effects, which are studied in detail by earlier authors
\cite{Dighe00,Takahashi01} (MSW-H at $\sim 0.05 R_\odot$, MSW-L at $\sim
(0.2-0.5) R_\odot$).
The strange behavior of $P(\nu_\mu \rightarrow \bar{\nu}_e)$ at $r
\sim0.04 R_\odot$ is considered due to the nonadiabatic RSF-L conversion,
while the probability is not changed as a whole.

\begin{figure}[htbp]
\includegraphics[width=15cm]{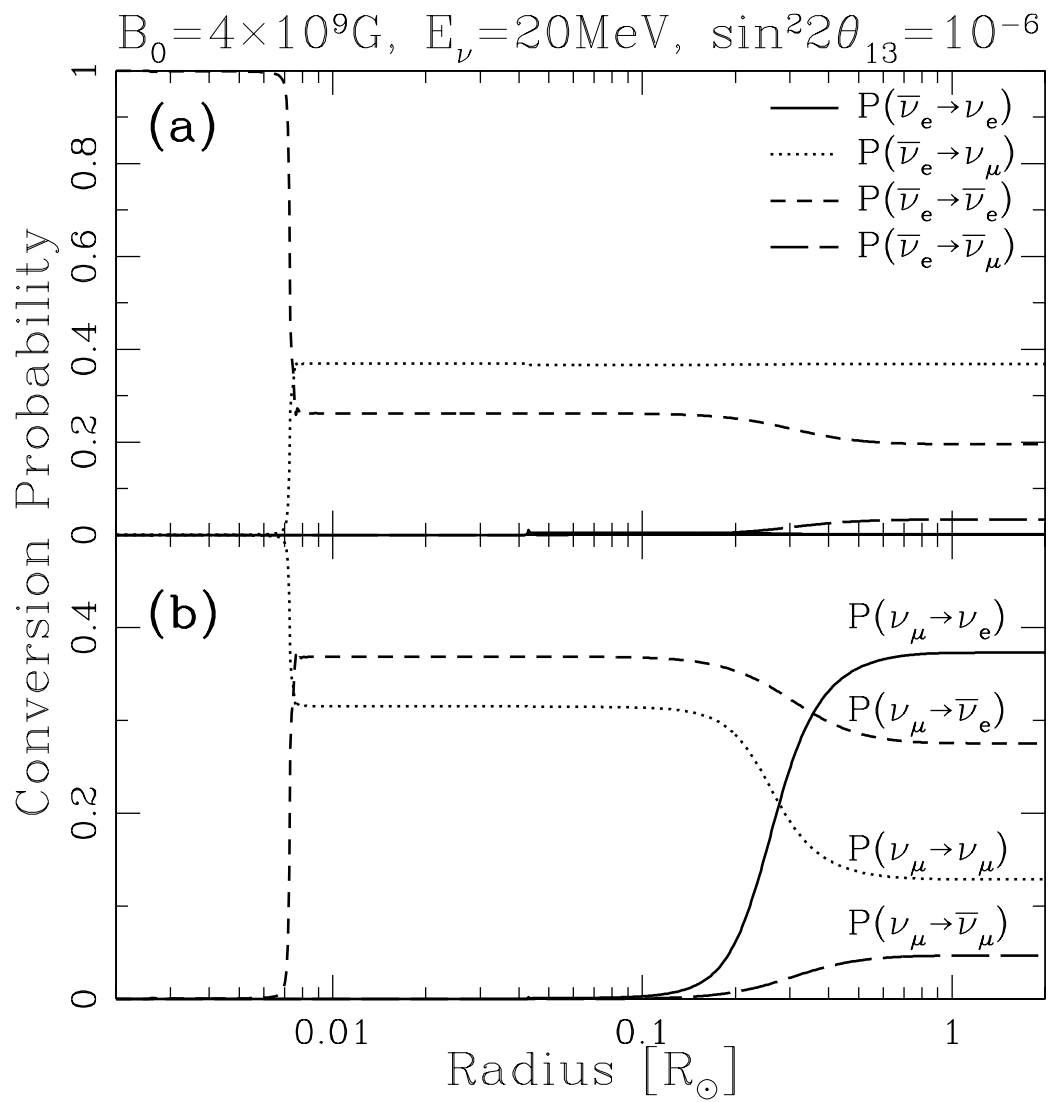}%
\caption{The same figure as Fig. \ref{fig:prob_L}, but for $\sin^2 
 2 \theta_{13} = 10^{-6}$. \label{fig:prob_S}}
\end{figure}

Figure \ref{fig:prob_S} is the same one as Fig. \ref{fig:prob_L}, but it
is for $\sin^2 2 \theta_{13} = 10^{-6}$ instead of $0.04$.
In this case, we do not see the MSW-H conversion, since it is
nonadiabatic.

\begin{figure}[htbp]
\includegraphics[width=15cm]{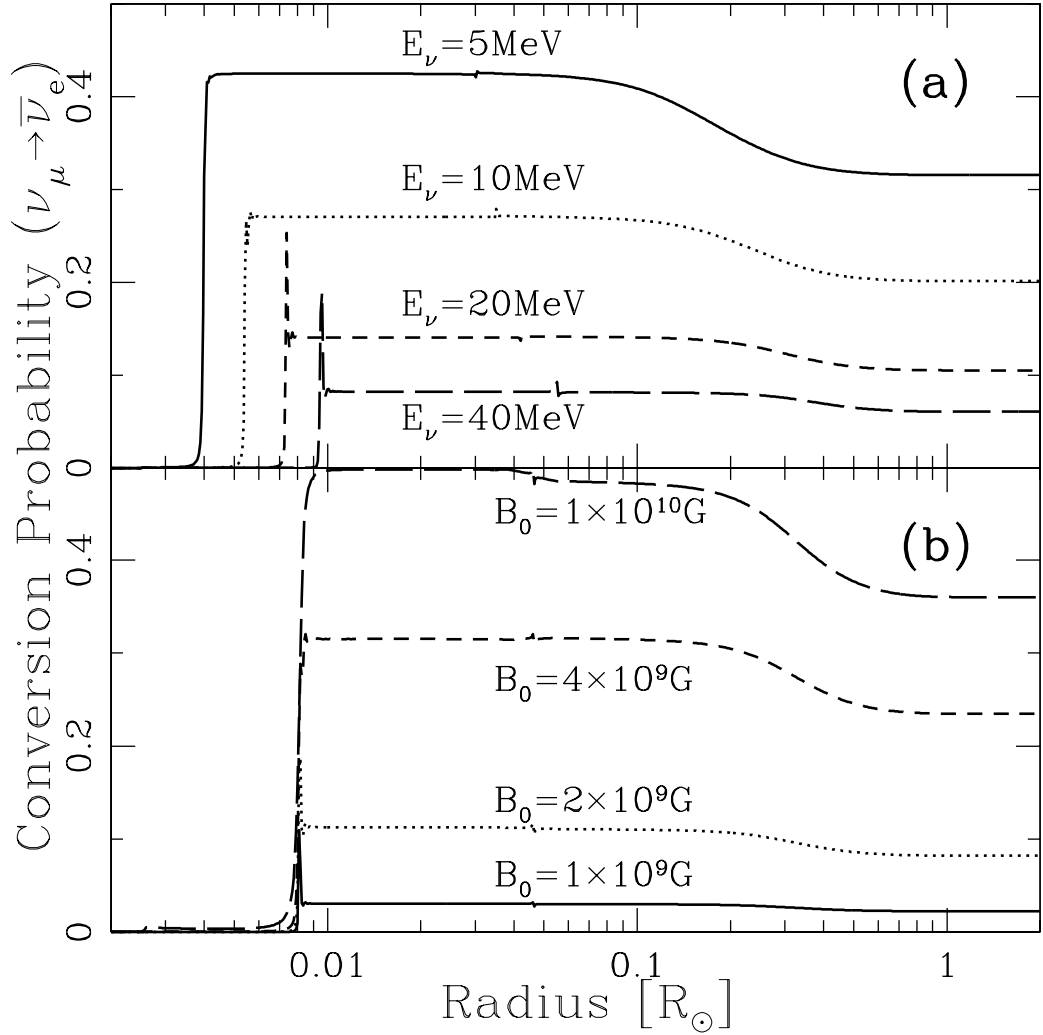}%
\caption{Conversion probability $P(\nu_\mu \rightarrow \bar{\nu}_e)$
 for $\sin^2 2 \theta_{13} = 0.04$. (a) The energy dependence for $B_0 = 
 2 \times 10^9 ~\mathrm G$. (b) The dependence on the magnetic field
 strength at the iron core, for $E_\nu = 25~\mathrm{MeV}$.
 \label{fig:prob_em}}
\end{figure}

We show in Fig. \ref{fig:prob_em}, the dependence of the conversion
probability $P(\nu_\mu \rightarrow \bar{\nu}_e)$ on the energy and the
magnetic field strength at the iron core, assuming $\sin^2 2 \theta_{13}
= 0.04$.
From Fig. \ref{fig:prob_em} (a), we see that the resonance point goes to
outer region and the conversion becomes more and more nonadiabatic, as
the neutrino energy becomes larger.
In the same manner, Fig. \ref{fig:prob_em} (b) indicates that the
conversion becomes more adiabatic for larger magnetic field strength.
These behaviors can be understood qualitatively using
Fig. \ref{fig:profiles} and Eq. (\ref{eq:adiabaticity}).

\begin{figure}[htbp]
\includegraphics[width=15cm]{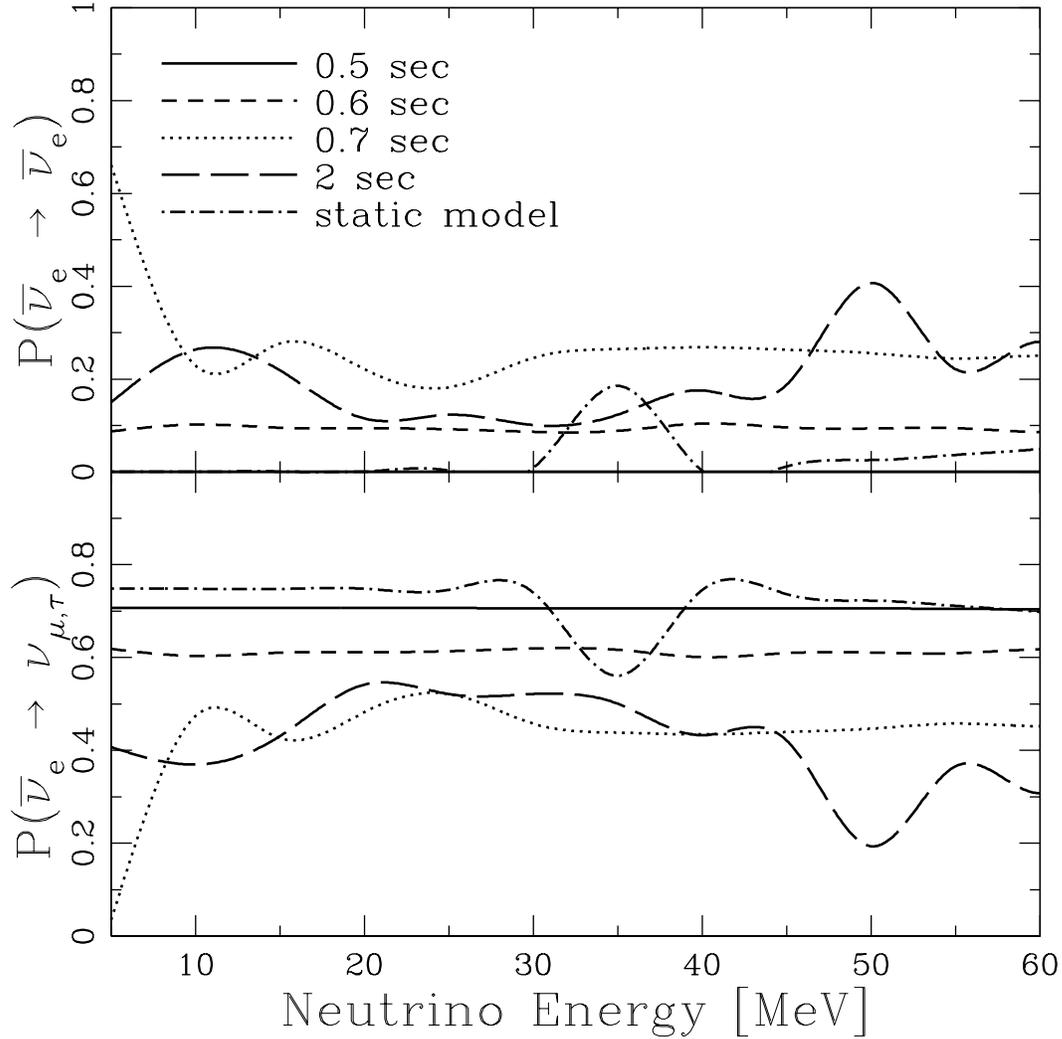}
\caption{The conversion probability $P(\bar{\nu}_e \to \bar{\nu}_e)$ and 
 $P(\bar{\nu}_e \to \nu_{\mu,\tau})$ as a function of neutrino energy,
 which are evaluated with time-dependent density and $Y_e$ profiles at
 0.5 (solid line), 0.6 (short-dashed line), 0.7 (dotted line), and 2
 (long-dashed line) seconds after bounce. Those obtained with static
 progenitor model is also shown as dot-dashed line. The magnetic field
 profile is assumed to be the same one as the static dipole model ($B_0
 = 10^{10}$ G). \label{fig:conv_prob}}
\end{figure}

At the end of this subsection, we compare the calculated conversion
probabilities with ones which are evaluated with time-dependent density
profile.
Figure \ref{fig:conv_prob} shows $P(\bar{\nu}_e \to \bar{\nu}_e)$ and
$P(\bar{\nu}_e \to \nu_{\mu,\tau})$ as a function of neutrino energy.
Since how the shock propagation changes the structure of magnetic field
at resonance points is not known well, we assume the same static dipole
profile as for magnetic field, and the figure is for $B_0=10^{10}$ G.
As discussed in Section \ref{sub:Supernova magnetic field}, this
assumption is considered to be a good one for $\alt 0.5$ seconds after
bounce, however, probably incorrect for $>0.5$ seconds.
Anyway, from the figure, we can confirm that the probabilities are in
good agreement between the model 0.5 sec and the static progenitor
model.

\subsection{Energy spectrum at the SuperKamiokande detector
\label{sub:Energy spectrum at the SuperKamiokande}}

SK is a water \v{C}herenkov detector with 32,000 tons of pure water
based at Kamioka in Japan. The relevant interactions of neutrinos with
water are as follows:
\begin{eqnarray}
\bar{\nu}_e + p &\rightarrow& e^+ + n, ~\mbox{(C.C.)} 
\label{eq:nuebar+p} \\
\nu_e + e^- &\rightarrow& \nu_e + e^-, ~\mbox{(C.C. and N.C.)} 
\label{eq:nue+e} \\
\bar{\nu}_e + e^- &\rightarrow& \bar{\nu}_e + e^-, ~\mbox{(C.C. and N.C.)}
\label{eq:nuebar+e} \\
\nu_x + e^- &\rightarrow& \nu_x + e^-, ~\mbox{(N.C.)}
\label{eq:nux+e} \\
\nu_e + \mbox{O} &\rightarrow& e^- + \mbox{F}, ~\mbox{(C.C.)}
\label{eq:nue+O} \\
\bar{\nu}_e + \mbox{O} &\rightarrow& e^+ + \mbox{N}, ~\mbox{(C.C.)}
\label{eq:nuebar+O}
\end{eqnarray}
where C.C. and N.C. stand for charged- and neutral-current interactions,
respectively.

SK is not in operation now because of the unfortunate accident, which
occurred on 12th November 2001.
However, it is expected to be repaired using remaining photomultiplier
tubes (PMTs), and will restart the detection by the end of this year
(2002) with lower performance.
The effect of the accident on its performance is expected not to be
serious for supernova neutrinos, because the fiducial volume will not
change, and the threshold energy change (from 5 MeV to about 7-8 MeV)
influences the event number very little. 
Although the energy resolution will become about $\sqrt{2}$ times worse,
it does not matter for our consideration.
In the calculations, we used the energy threshold and the energy
resolution after the accident.

For the reaction cross sections, we refer to Ref. \cite{Totsuka92}.
The $\bar{\nu}_e p$ is the largest reaction and the energy of recoil
positrons is related to that of injected neutrinos by simple formula
$E_e = E_{\bar{\nu}_e} - (M_n - M_p) = E_{\bar{\nu}_e} -
1.29~\mathrm{MeV}$, where $M_n,~M_p$ are neutron and proton mass,
respectively.
The energy spectrum of electrons (positrons), therefore, reflects that
of $\bar{\nu}_e$, which is easily estimated in some simple cases using
e.g., Eqs. (\ref{eq:F_ebar_A_NA}), (\ref{eq:F_ebar_NA_NA}),
(\ref{eq:F_ebar_A_A}), and (\ref{eq:F_ebar_NA_A}).

\begin{figure}[htbp]
\includegraphics[width=15cm]{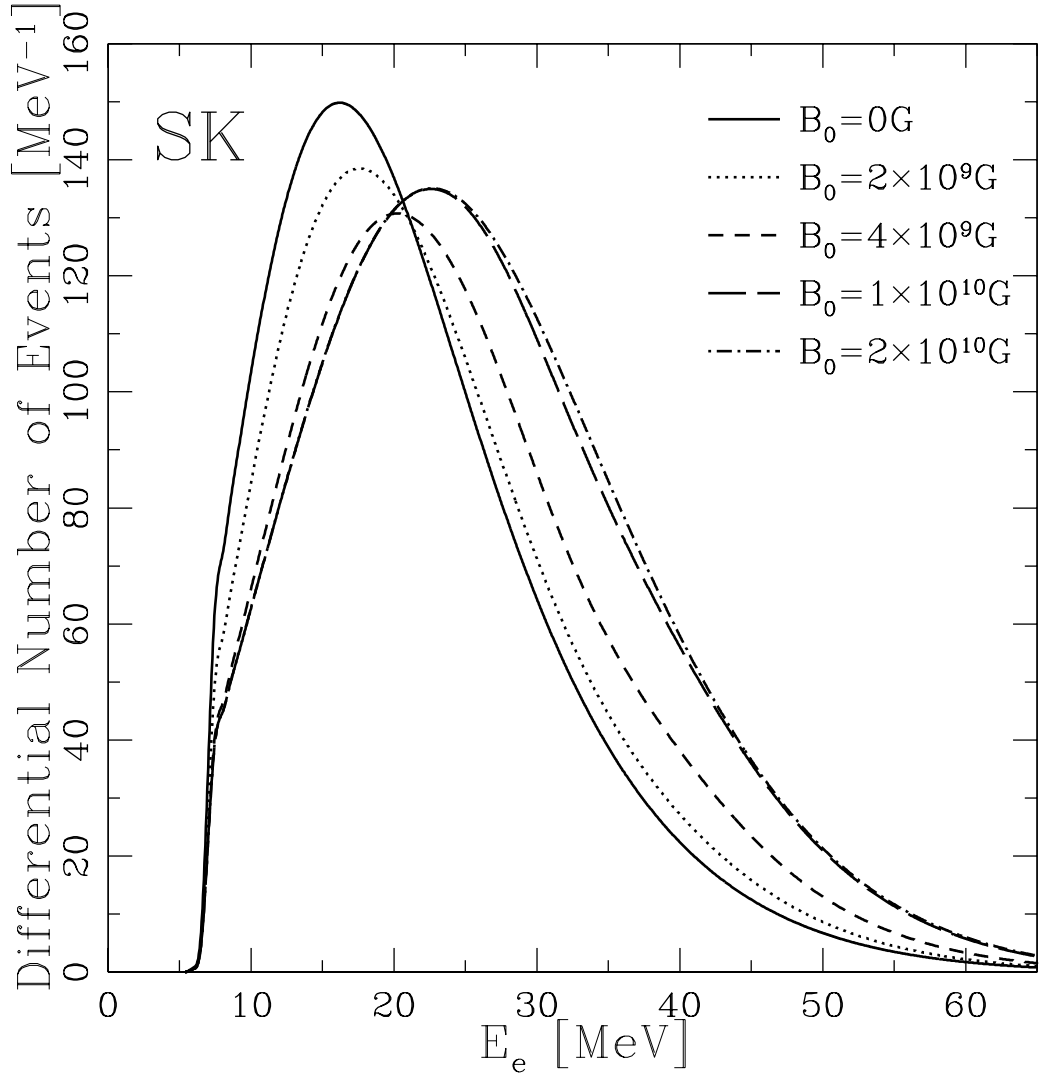}%
\caption{Energy spectra of electrons (positrons) at SK for first 0.5
 seconds, evaluated with $\sin^2 2 \theta_{13} = 0.04$ and several
 values of  $B_0$. The energy spectra for $\sin^2 2 \theta_{13} =
 10^{-6}$ are almost the same. The mean energy increases as $B_0$
 becomes larger, and saturates around $B_0 = 10^{10} ~\mathrm G$
 (completely adiabatic RSF-H conversion). \label{fig:spectra_SK}}
\end{figure}

Figure \ref{fig:spectra_SK} shows the energy spectra of electrons
(positrons) for first 0.5 seconds evaluated with $\sin^2 2 \theta_{13} =
0.04$ and several values of $B_0$, which are calculated with conversion
probabilities, original neutrino spectra, reaction cross sections, and
detector performance.
When we increase the value of $B_0$, the energy spectrum shifts to
higher energy region and the shift saturates at $B_0 \sim
10^{10}~\mathrm G$.
This is because at $B_0 \agt 10^9 ~\mathrm G$ the RSF-H conversion
starts to become adiabatic and it achieves completely adiabatic
transition at $B_0 \sim 10^{10} ~\mathrm G$. 
For $\sin^2 2 \theta_{13} = 10^{-6}$, we obtained almost the same
figure as Fig. \ref{fig:spectra_SK}, and we do not show it here.

\subsection{Energy spectrum at the Sudbury Neutrino Observatory
\label{sub:Energy spectrum at the Sudbury Neutrino Observatory}}

SNO is a \v{C}herenkov detector filled with 1,000 tons of heavy water.
The interactions of neutrinos with heavy water are 
\begin{eqnarray}
\nu_e + d &\rightarrow& e^- + p + p, ~\mbox{(C.C.)}
\label{eq:nue+d} \\
\bar{\nu}_e + d &\rightarrow& e^+ + n + n, ~\mbox{(C.C.)}
\label{eq:nuebar+d} \\
\nu_x + d &\rightarrow& \nu_x + p + n, ~\mbox{(N.C.)}
\label{eq:nux+d} 
\end{eqnarray}
and reactions represented by Eqs. (\ref{eq:nue+e}) to (\ref{eq:nuebar+O}).
Furthermore, the produced neutrons are detected through the interaction
with surrounding nuclei as
\begin{eqnarray}
n + d &\rightarrow& {}^3\mathrm H + \gamma_{6.3}, 
~(\mbox{efficiency: 24\%})
\label{eq:n+d} \\
n + {}^{35}\mathrm{Cl} &\rightarrow& {}^{36}\mathrm{Cl}  + \gamma_{8.6},
~(\mbox{efficiency: 83\%})
\label{eq:n+Cl}
\end{eqnarray}
where $\gamma_{6.3}$ and $\gamma_{8.6}$ represent the
$\gamma$-rays whose energies are 6.3 MeV and 8.6 MeV, respectively, and
NaCl is added to efficiently capture neutrons.
Owing to these neutron capture reactions, the neutrino detection through
the neutral-current reaction Eq. (\ref{eq:nux+d}) is possible.
Furthermore from the charged-current reaction of $\bar{\nu}_e$'s
(Eq. (\ref{eq:nuebar+d})), we receive more than two signals, i.e., one
from the preceding recoil positron, and the others from the delayed
$\gamma$-rays by neutron captures.
Thus, using this criterion, we can discern the $\bar{\nu}_e d$ events
from the other detection channels.

\begin{figure}[htbp]
\includegraphics[width=15cm]{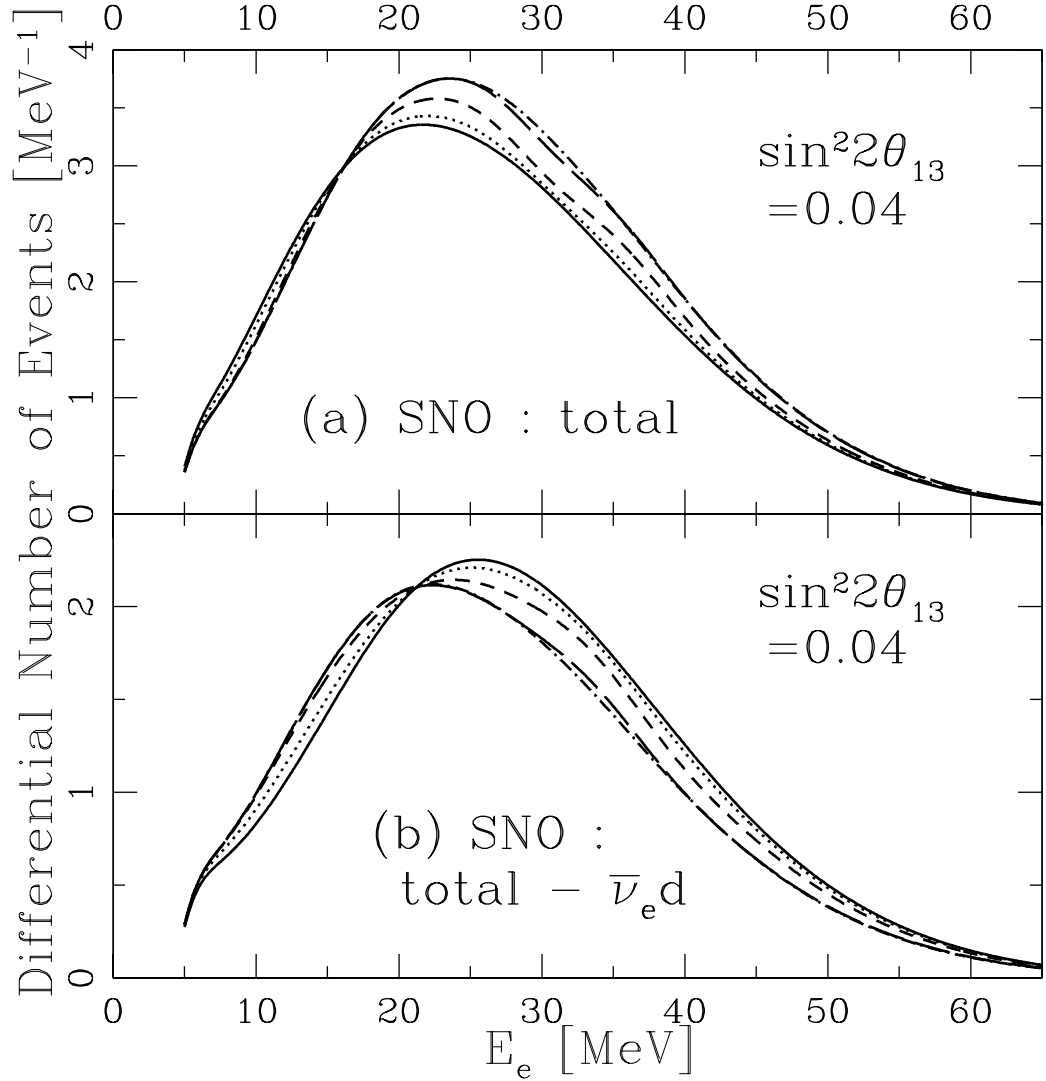}%
\caption{Energy spectra at SNO for first 0.5 seconds evaluated with
 $\sin^2 2 \theta_{13} = 0.04$ and several values of $B_0$. The line
 types are the same as those used in Fig. \ref{fig:spectra_SK}. (a)
 Total numbers of events at SNO. As $B_0$  increases, the spectrum
 becomes harder. (b) Numbers of events at SNO, where the $\bar{\nu}_e d$
 events (Eq. (\ref{eq:nuebar+d})) are subtracted from the total
 events. The spectrum moves conversely compared to the total events (a),
 as the values of $B_0$. \label{fig:spectra_SNO_L}}
\end{figure}

For the cross sections of reactions with deuterons, we referred to
Ref. \cite{Ying89}.
Figure \ref{fig:spectra_SNO_L} shows the energy spectra for first 0.5
seconds evaluated with $\sin^2 2 \theta_{13} = 0.04$ and several values
of $B_0$ \footnote{Since the neutral-current reaction
Eq. (\ref{eq:nux+d}) does not contribute to energy spectrum, we neglect
it from here on.}.
The upper panel (a) shows the total energy spectra and the lower panel
(b) the spectra of (total$-\bar{\nu}_e d$) events which can be obtained
using the delayed coincidence signals, where the line types are the same
as those used in Fig. \ref{fig:spectra_SK}.
As $B_0$ inceases, the energy spectrum of the total events becomes
harder, while that of the (total$- \bar{\nu}_e d$) events softer.
These properties can be explained as follows.
Since almost all the events come from $\nu_e d$ and $\bar{\nu}_e d$
reactions, the (total$-\bar{\nu}_e d$) events are roughly coincide with
$\nu_e d$ events.
Thus, from the qualitative discussion given in Section
\ref{sec:Level crossing schemes and qualitative illustrations}, we can
obtain the properties in Fig. \ref{fig:spectra_SNO_L} (b) (see
Eqs. (\ref{eq:F_e_A_NA}) and (\ref{eq:F_e_A_A})).
Since the energy spectrum due to $\bar{\nu}_e d$ becomes harder for
larger $B_0$ (see Eqs. (\ref{eq:F_ebar_A_NA}) and
(\ref{eq:F_ebar_A_A})), the total spectrum also becomes harder but with
the suppressed degree of the shift, because that of the $\nu_e d$ events
relaxes.

\begin{figure}[htbp]
\includegraphics[width=15cm]{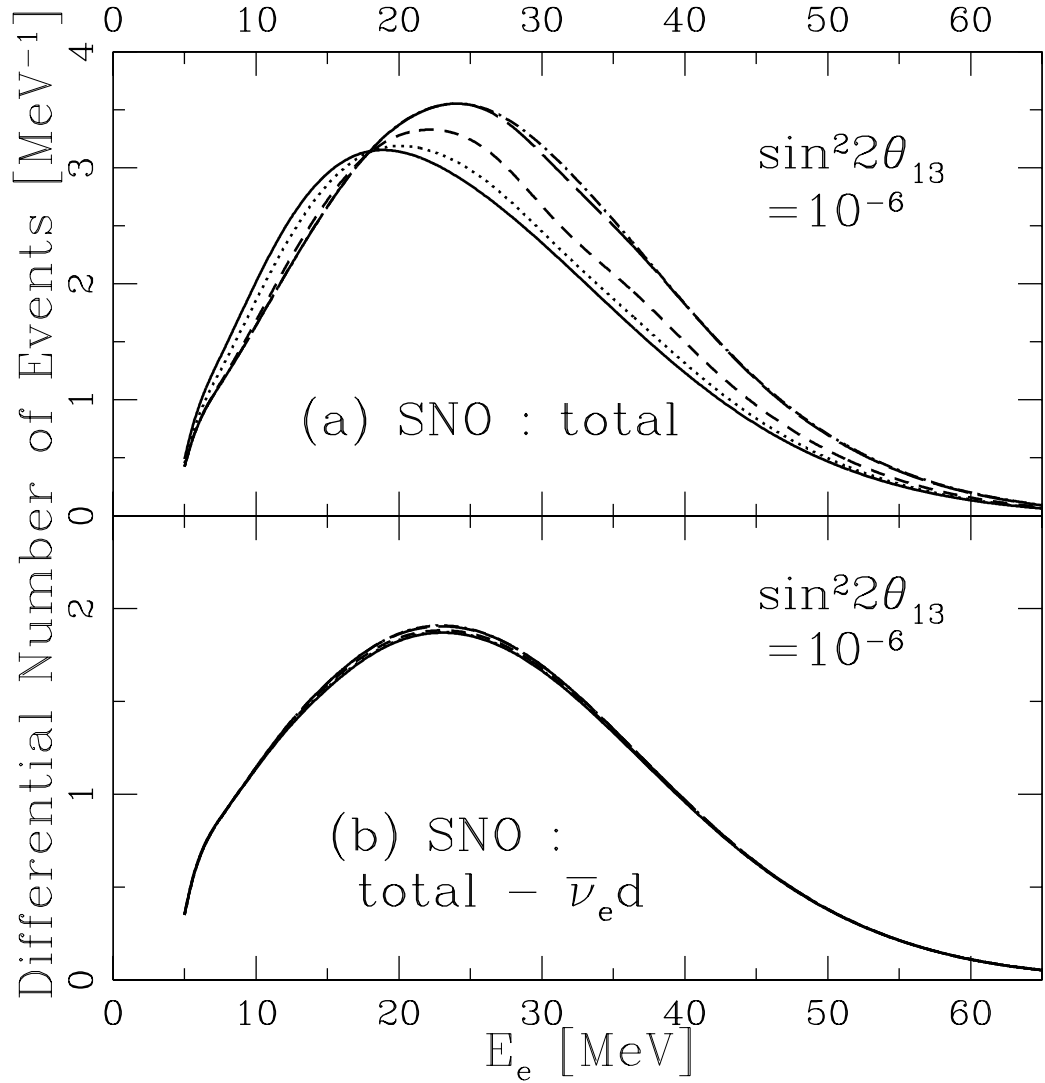}%
\caption{The same figure as Fig. \ref{fig:spectra_SNO_L}, but for
 $\sin^2 2 \theta_{13} = 10^{-6}$. For larger $B_0$, the spectrum of the
 total events (a) becomes harder, while that of (total$- \bar{\nu}_e d$)
 events (b) changes little. \label{fig:spectra_SNO_S}}
\end{figure}

Figure \ref{fig:spectra_SNO_S} is the same one as
Fig. \ref{fig:spectra_SNO_L}, but for the different value of
$\theta_{13}$ ($\sin^2 2 \theta_{13} = 10^{-6}$).
For larger $B_0$, the spectrum of the total events becomes harder, while
that of (total$- \bar{\nu}_e d$) events changes little (see
Eqs. (\ref{eq:F_e_NA_NA}), (\ref{eq:F_ebar_NA_NA}), (\ref{eq:F_e_NA_A}),
and (\ref{eq:F_ebar_NA_A})).

\section{Discussion \label{sec:Discussion}}

In this study, we used the original neutrino spectrum calculated by
Lawrence Livermore group \cite{Totani98}.
Unfortunately, their study as well as the other published full numerical
supernova collapse simulations have not yet included the nucleon
bremsstrahlung process or nucleon recoils, even though it is no longer
controversial that these effects are important.
Recent studies (e.g. \cite{Keil02}) including all these processes have
shown that average $\nu_x$ energy exceeds the average $\bar{\nu}_e$
energy by only a small amount, 10\% being a typical number.
If this is the case, the spectral deformation, particularly at SK
detector, will be little and it is difficult to obtain some information
on the flavor conversion mechanism in supernova.
However, it is premature to conclude that their results are correct,
since it is based on the neutrino transport study on the background of
an assumed neutron star atmosphere, and this approach lacks
hydro-dynamical self-consistency.
Further it is also because the mean energies and their ratios change
significantly between the supernova bounce, accretion phase, and the
later neutron star cooling phase.

Thus, at present, even if the Galactic supernova neutrino burst were
detected by SK and SNO, it would be very difficult to obtain useful
information on whether the RSF effect actually occurs in supernova and
to obtain implication for neutrino magnetic moment, owing to large
uncertainties concerning original neutrino spectrum.
In the future, however, it might be plausible that these uncertainties
will be removed by development of numerical simulations, and the signal
of supernova neutrino burst would provide valuable implications for the
flavor conversion mechanisms.
In discussions from here, we present the method to probe conversion
mechanisms from the detected neutrino signals, {\it assuming} that the
uncertainties concerning the supernova models are considerably reduced
by the future development of numerical approaches.

\subsection{Implication for neutrino magnetic moment \label{sub:Implication on neutrino magnetic moment}}

As we have shown in Section \ref{sec:Results}, the deformation of
spectral shape occurs when $B_0 \agt 10^9 ~\mathrm G$.
Therefore the ratio of high energy events to low energy ones is a good
measure to investigate the magnetic field strength in supernova or
neutrino magnetic moment.
We adopted as that two quantities below:
\begin{eqnarray}
R_{\mathrm{SK}} &=& \frac{\mbox{Number of events for }E_e > 25 ~\mathrm{MeV}}
 {\mbox{Number of events for }E_e < 20 ~\mathrm{MeV}}, \\
R_{\mathrm{SNO}} &=& \frac{\mbox{Number of events for }E_e > 25 ~\mathrm{MeV}}
 {\mbox{Number of events for }E_e < 15 ~\mathrm{MeV}}.
\end{eqnarray}

\begin{figure}[htbp]
\includegraphics[width=15cm]{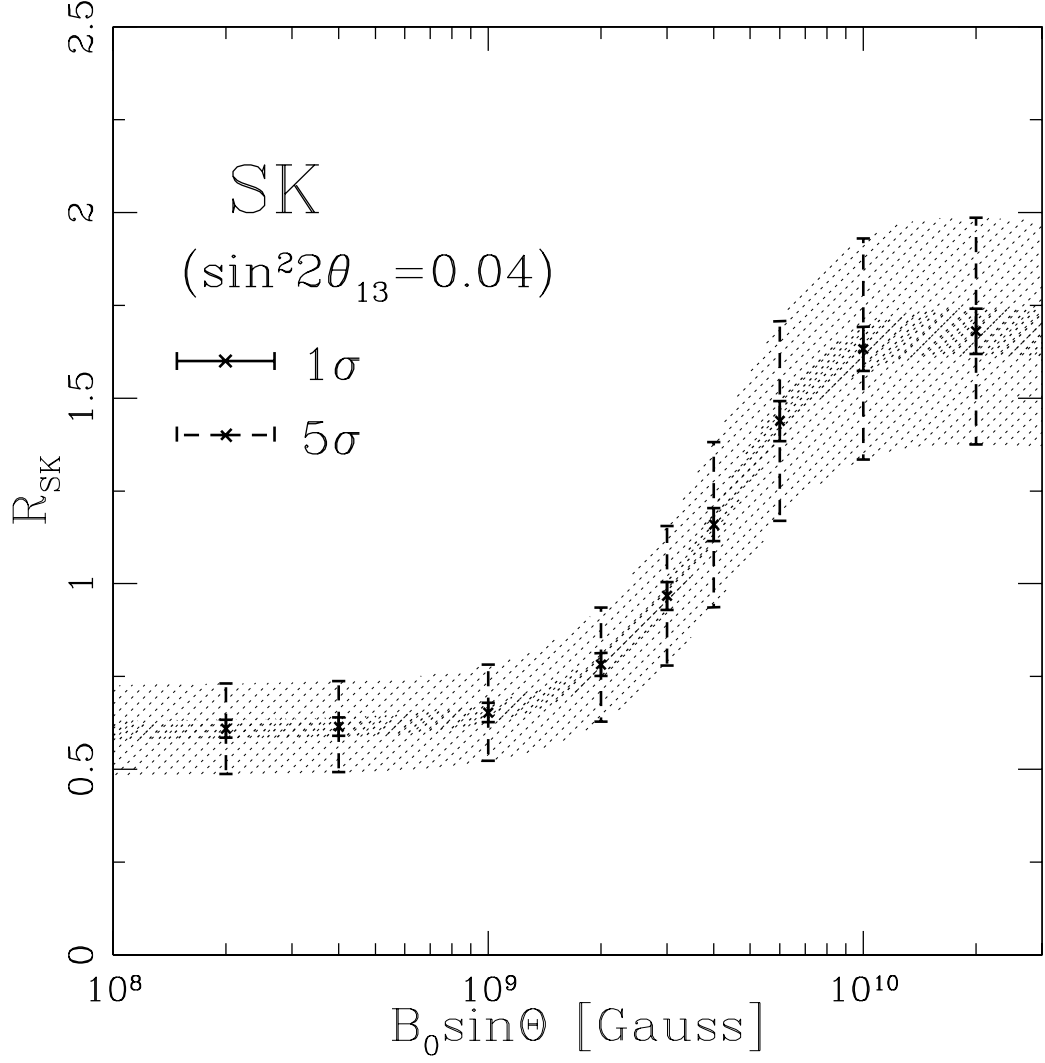}%
\caption{Ratio of high energy events to low energy events at SK,
 $R_{\mathrm{SK}}$, as a function of $B_0 \sin \Theta$ for $\sin^2 2
 \theta_{13} = 0.04$, where error bars include only statistical errors.
 The solid error bars are at $1 \sigma$ level and the dashed ones at $5
 \sigma$ level. The thick and thin bands are the cubic spline
 interpolations of $1 \sigma$ and $5 \sigma$ error bars,
 respectively. \label{fig:ratio_SK}}
\end{figure}

Figure \ref{fig:ratio_SK} shows $R_{\mathrm{SK}}$ as a function of
$B_0 \sin \Theta$ for $\sin^2 2 \theta_{13} = 0.04$, where error bars
include only statistical errors.
(Here, we re-introduce the angular factor $\sin \Theta$ since we also
discuss the dependence on magnetic field orientation below.)
We obtained almost the same figure for $\sin^2 2 \theta_{13} = 10^{-6}$.
Therefore, SK data will provide statistically sufficient information on
the magnetic field strength in the supernova and neutrino magnetic
moment, and these consequences are independent of $\theta_{13}$.
For example, the value of $R_{\mathrm{SK}}$ which is found to be larger
than $\sim 1$ indicates that the RSF conversion occurred in the
supernova, and we can constrain the value of $\mu_\nu B_0 \sin \Theta$
assuming the supernova and presupernova models.

\begin{figure}[htbp]
\includegraphics[width=15cm]{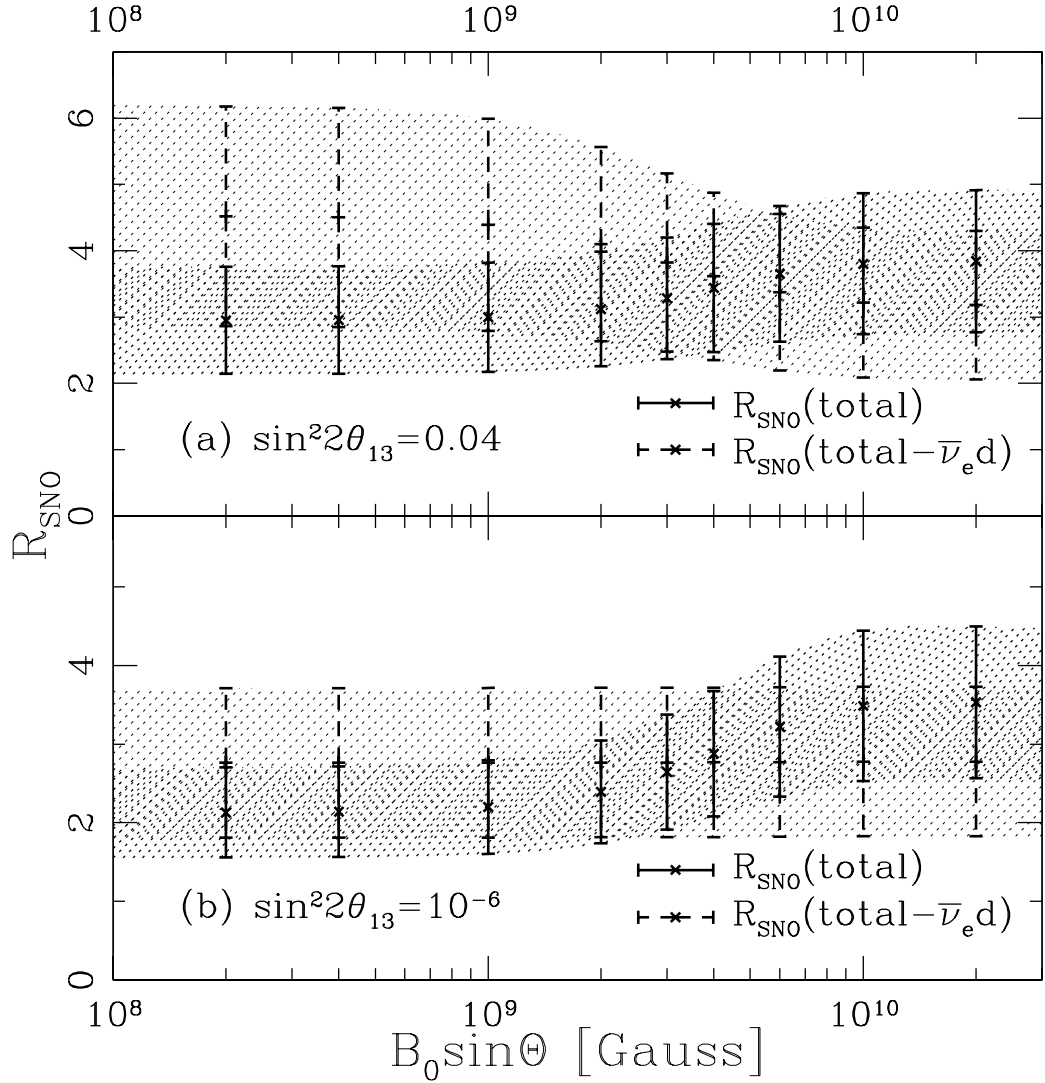}%
\caption{Ratio of high energy events to low energy events at SNO,
 $R_{\mathrm{SNO}}$, where the upper panel (a) is for $\sin^2 2
 \theta_{13} = 0.04$ and the lower panel (b) for $\sin^2 2 \theta_{13}
 = 10^{-6}$. The error bars include only statistical errors, and are
 at $1 \sigma$ level. The thick and thin bands are the cubic spline
 interpolations of these error bars. \label{fig:ratio_SNO}}
\end{figure}

The same figure for SNO is Fig. \ref{fig:ratio_SNO}, where the upper
panel (a) is for $\sin^2 2 \theta_{13} = 0.04$ and the lower panel (b)
for $\sin^2 2 \theta_{13} = 10^{-6}$.
If we obtain $R_{\mathrm{SNO}}(\mathrm{total})$ which is larger than
$R_{\mathrm{SNO}}(\mathrm{total - \bar{\nu}_e d})$, it suggests that
the supernova magnetic field is quite strong, $B_0 \sin \Theta \agt
5 \times 10^9 \mathrm G$, and neutrinos have non-zero magnetic moment.
However, since the statistical errors of $R_{\mathrm{SNO}}$ are rather
large because of smallness of data, we can use the SNO data to confirm
the SK result.

In addition, if we know the value of the angular factor $\sin \Theta$
from the optical observation such as the Hubble Space Telescope (HST),
we can constrain the magnetic field strength $B_0$ (not the combination
$B_0 \sin \Theta$) from the neutrino observation.
Actually, in the case of SN1987A, which occurred in the Large Magellanic
Cloud (LMC), $\Theta$ is constrained with the procedure as follows. 
The HST images of SN1987A show a bright elliptical ring surrounding the
optical remnant.
This inner ring has been interpreted as a ring of high-density material
in the equatorial plane of the progenitor, caused by the impact of the
high velocity stellar wind of the blue supergiant progenitor with its
earlier, slower red giant wind, and ionized by the UV flash from the
supernova explosion \cite{Burrows95}. 
The inner ring is known to be inclined to the line of sight by $\sim
43^\circ$.
Thus, assuming that the magnetic dipole moment of the star orient to the
rotation axis, we can constrain $\sin \Theta$ to be $\sim 1 / \sqrt{2}$
from the HST observation.
However, if the next supernova occurred in the Galactic center, it would
be quite difficult to estimate $\Theta$ since the Galactic center is
optically thick from the Earth.
On the other hand, if it occurred in the optically thin environment such
as LMC, we would be able to constrain $\Theta$ from optical observations
and $B_0$ from the neutrino signals.
Although LMC is located at $\sim 50$ kpc away from the Earth, SK data
would permit statistically sufficient discussions on the supernova
magnetic field.
For example, in that case, the $1 \sigma$ statistical errors of
$R_{\mathrm{SK}}$ are as large as $5 \sigma$ errors in the case of the
Galactic supernova ($D = 10 ~\mathrm{kpc}$), which are also shown in
Fig. \ref{fig:ratio_SK}, and it will be sufficient to distinguish
whether $B_0 \sin \Theta \agt 5 \times 10^9 ~\mathrm G$ or $B_0 \sin
\Theta \alt 10^9 ~\mathrm G$.

\subsection{Constraint from supernova relic neutrino observation
\label{sub:Constraint from supernova relic neutrino observation}}

In addition to next Galactic supernova, the observation of supernova
relic neutrino (SRN) background would be available to obtain information
on flavor conversion mechanism in supernova.
The flux of SRN is calculated by
\begin{equation}
\frac{dF_\nu}{dE_\nu} = c \int^{z_\mathrm{max}}_0 R_{\mathrm{SN}}(z)
 \frac{dN_\nu(E_\nu^\prime)}{dE_\nu^\prime} (1+z) \frac{dt}{dz} dz,
\end{equation}
where $E_\nu^\prime = (1+z) E_\nu$, $R_{\mathrm{SN}}(z)$ is supernova
rate per comoving volume at redshift $z$, $dN_\nu / dE_\nu$ energy
spectrum of emitted neutrinos, and $z_{\mathrm{max}}$ the redshift when
the gravitational collapses began (see Ref. \cite{Ando02} for detailed
discussion).

Most recently, SK collaboration set an upper bound of $1.2 ~\bar{\nu}_e
\mathrm{~cm^{-2}~s^{-1}}$ for the SRN flux in the energy region $E_\nu >
19.3$ MeV \cite{Malek02}.
That limit is the same order as some typical theoretical predictions
such as Ref. \cite{Ando02}.
For example, Ando et al. \cite{Ando02} predicted that the total SRN flux
integrated over entire energy was $11 \mathrm{~cm^{-2}~s^{-1}}$, while
the corresponding SK limit calculated with the predicted spectral shape
is $31 \mathrm{~cm^{-2}~s^{-1}}$.
Since the theoretical calculations contain many ambiguities such as the
supernova rate in the universe and neutrino spectrum from each
supernova, this severe observational SRN limit can provide a number of
valuable information on the various fields of astrophysics and
cosmology.
Further, in the near future, it is expected that the upper limit will be
much lower (about factor 3) when the current SK data of 1,496 days are
reanalyzed using some technique to reduce background against detection
\cite{Suzuki02}.
In that case, the further severer constraint not only on the star
formation history but also on the nature of neutrinos might be obtained.

Since the RSF mechanism further enhances the average energy of
$\bar{\nu}_e$ compared to the ordinary MSW effect, the current and the
future expected upper limit on the SRN flux might provide very useful
information on that mechanism.
However, in order to give precise SRN prediction including the RSF
mechanism, we must calculate conversion probability in supernova for
full time-scale during the neutrino burst.
Therefore, the prediction is difficult at present stage, and it is also
one of our future works.

\section{Conclusion \label{sec:Conclusion}}

We investigated three-flavor neutrino oscillation with non-zero neutrino
magnetic moment in supernova magnetic fields.
We gave a new crossing diagram which includes not only ordinary MSW
resonance but also magnetically-induced RSF effect.
From the diagram, it was found that there occurred four resonances, two
were induced by the RSF effect and two by the pure MSW.
In addition, the qualitative behavior of neutrino conversions and
expected neutrino spectrum at the detector were found to be well
illustrated.

We also numerically calculated conversion probability in the supernova
to obtain the expected energy spectrum at the detector.
In that calculation, realistic neutrino mixing parameters inferred by
recent neutrino oscillation experiments and realistic density and $Y_e$
profiles of presupernova stars were adopted.
As for magnetic field in the supernova, we assumed the dipole structure,
and that the field strength was normalized at the surface of the iron
core, which was inferred from the observations of white dwarfs.
Although the shock wave propagation changes density profile in the
resonance region drastically and is also expected to change magnetic
filed profile at the same region, we showed that if we confine our
discussion until 0.5 seconds after core bounce, using static progenitor
model and static magnetic field structure is expected to be a good
approximation.

As a result of numerical calculations, it was found that if the magnetic
field at the surface of the iron core is sufficiently strong ($B_0 \sim
10^{10} \mathrm G$, where we assumed $\mu_\nu = 10^{-12} ~\mu_B$), the
RSF-H conversion becomes adiabatic and complete $\bar{\nu}_e
\leftrightarrow \nu_\tau$ conversion occurs.
Hence, non-zero neutrino magnetic moment with the strong magnetic field
enhances the energies of $\bar{\nu}_e$ and hardens the energy spectrum at
SK and SNO.
Thus, in the future, if the uncertainty concerning original neutrino
spectrum is removed by development of numerical simulations for
supernova explosions, we might be able to obtain information on neutrino
magnetic moment as well as the magnetic field strength in supernova
using the energy spectrum at detectors.
In addition, it could be also plausible using the current and the future
supernova relic neutrino observations.

% Surround figure environment with turnpage environment for landscape
% figure
% \begin{turnpage}
% \begin{figure}
% \includegraphics{}%
% \caption{\label{}}
% \end{figure}
% \end{turnpage}

% tables should appear as floats within the text
%
% Here is an example of the general form of a table:
% Fill in the caption in the braces of the \caption{} command. Put the label
% that you will use with \ref{} command in the braces of the \label{} command.
% Insert the column specifiers (l, r, c, d, etc.) in the empty braces of the
% \begin{tabular}{} command.
% The ruledtabular environment adds doubled rules to table and sets a
% reasonable default table settings.
% Use the table* environment to get a full-width table in two-column
% Add \usepackage{longtable} and the longtable (or longtable*}
% environment for nicely formatted long tables. Or use the the [H]
% placement option to break a long table (with less control than 
% in longtable).
% \begin{table}%[H] add [H] placement to break table across pages
% \caption{\label{}}
% \begin{ruledtabular}
% \begin{tabular}{}
% Lines of table here ending with \\
% \end{tabular}
% \end{ruledtabular}
% \end{table}

% Surround table environment with turnpage environment for landscape
% table
% \begin{turnpage}
% \begin{table}
% \caption{\label{}}
% \begin{ruledtabular}
% \begin{tabular}{}
% \end{tabular}
% \end{ruledtabular}
% \end{table}
% \end{turnpage}

% Specify following sections are appendices. Use \appendix* if there
% only one appendix.
%\appendix
%\section{}

% If you have acknowledgments, this puts in the proper section head.
\begin{acknowledgments}
% put your acknowledgments here.
We would like to thank SuperKamiokande collaboration including
 Y. Suzuki, M. Nakahata, Y. Fukuda, and A. Suzuki for useful discussion
 on supernova relic neutrino background.
This work was supported in part by grants-in-aid for scientific research
 provided by the Ministry of Education, Science and Culture of Japan
 through Research grant no. S14102004.
\end{acknowledgments}

% Create the reference section using BibTeX:
%\bibliography{paper.bib}

\clearpage

\end{document}